\documentclass[aps,preprint,prd]{revtex4}

\usepackage[english]{babel}
\usepackage[utf8x]{inputenc}
\usepackage[T1]{fontenc}

\usepackage[a4paper,top=3cm,bottom=2cm,left=2cm,right=2cm,marginparwidth=1.75cm]{geometry}

\usepackage{amsmath}
\usepackage{graphicx}
\usepackage[colorinlistoftodos]{todonotes}
\usepackage[colorlinks=true, allcolors=blue]{hyperref}
\begin{document}
\title{Implementing the inverse type-II seesaw mechanism into the 3-3-1 model}
\author{Carlos Ant\^onio de Sousa Pires $^{a}$, Felipe Ferreira de Freitas$^{b}$, Jing Shu$^{b,c,d,e}$, Li Huang$^{b,c,f}$ and Pablo Wagner Vasconcelos Olegário$^{a}$}

%%%%%%%%
\affiliation{{a: Departamento de Física, Universidade Federal da Paraíba, Caixa Postal 5008, 58051-970,
Jo\~ao Pessoa, PB, Brazil},\\
{b: CAS Key Laboratory of Theoretical Physics, Institute of Theoretical Physics, Chinese Academy of Sciences, Beijing 100190, P. R. China},\\
{c: School of Physical Sciences, University of Chinese Academy of Sciences, Beijing 100049, P. R. China},\\
{d: CAS Center for Excellence in Particle Physics, Beijing 100049, China},\\
{e: Center for High Energy Physics, Peking University, Beijing 100871, China}, \\
{f: Laboratory for Elementary Particle Physics, Cornell University, Ithaca, NY 14853, USA}}

\begin{abstract}
After the LHC is turning on and accumulating more data, the TeV scale seesaw mechanisms for small neutrino masses in the form of inverse seesaw mechanisms are gaining more and more attention once they provide neutrino masses at sub-eV scale and  can be probed at the LHC.  Here we restrict our investigation to the inverse type II seesaw case  and implement it into the framework of the 3-3-1 model with right-handed neutrinos. As interesting result, the mechanism  provides small masses to both the standard neutrinos as well as to the right-handed ones. Its best signature  are the doubly charged scalars which are sextet by the 3-3-1 symmetry. We investigate their production at the LHC through the process $\sigma (p\,p \rightarrow Z^*, \gamma^* ,Z^{\prime} \rightarrow \Delta^{++}\,\Delta^{--})$ and their signal through four leptons final state decay channel.

\end{abstract}
\maketitle
\section{INTRODUCTION}

The end of second phase of the LHC and the future plans for the high luminosity LHC and the 100 TeV Chinese collider brings new encouraging perspectives for the theoretical physics community. These new perspectives usher the development of a plethora of models beyond Standard Model (BSM), each one with his particular signatures and phenomenological implications. An important class of BSM models are those that incorporate  seesaw mechanisms for small neutrino masses whose signature could be probed at current and future energies at the LHC. In this regard, the most popular mechanism capable of generating neutrino masses at sub-eV scale and providing a signature at TeV scale are the inverse seesaw mechanisms. The main idea of these mechanisms is that lepton number is  explicitly violated at very low energy scale.  

Inverse seesaw  mechanisms can be implemented into the standard model in three different ways. The inverse type I seesaw mechanism\cite{Mohapatra:1986bd} is, by far, the most well known. It consists in adding  to the standard model particle content new neutrinos in the singlet form. It is also possible to perform such mechanism by adding  triplet of scalars to the standard model content\cite{Li:1985hy,Lusignoli:1990yk,deS.Pires:2005au,Freitas:2014fda}. We refer to this case as the inverse type II seesaw mechanism. Another possibility of implementation  is by adding  triplet of fermions to the standard model content\cite{Ma:2009kh,Ibanez:2009du}. 

In this work we introduce a novel approach which consist on embedding the inverse type II seesaw mechanism into the structure of the 3-3-1 model with right-handed neutrinos\cite{Singer:1980sw,Montero:1992jk,Foot:1994ym} and explore their  phenomenological consequences. This new procedure requires the addition of a sextet of scalars to the original scalar content of the model. As main feature, this methodology provides small masses to both left and right-handed neutrinos.  For completeness, we develop the scalar sector of the model and focus our attention into the set of scalars which compose the sextet. We show that, after SU(3) symmetry breaking, these scalars  decouple from the original scalar content of the model. To demonstrate the viability of probing the new content presents in our model, we make use of Monte Carlo generators to test its   signature in the form of doubly charged scalar produced at the LHC. 

 The rest of the paper is organized as it follows: In the section \ref{sec_II} we revisit the different seesaw mechanisms and  in the section \ref{sec_III} we implement the Inverse Type II SeeSaw mechanism into the 3-3-1 model. We demonstrate the full particle spectrum and the new features that appear within this approach. In the section \ref{sec_IV} we present the possible manners to probe our model at the LHC and future colliders.  In the section \ref{sec_V} we present our conclusions.

\section{INVERSE SEESAW MECHANISMS}
\label{sec_II}
%%%%%%%
\subsection{Inverse type I seesaw mechanism}
%%%%
The implementation of the inverse type I seesaw mechanism\cite{Mohapatra:1986bd} into any particle physics model requires the  existence of six right-handed neutral fermions ($N_{i_R}\,,\,S_{i_R}$)  in addition to the three standard model ones $\nu_{i_L}$ with $i=1,2,3$. This mechanism requires the following mass terms,
\begin{equation}
{\cal L}=-\bar \nu_L m_D N_R  - \bar N_R M_N S^C_R- \frac{1}{2} \bar S_R \mu S_R^c + {\mbox h.c.}\,.
\label{massterms}
\end{equation}
With the basis $\nu =(\nu_L \,,\,N^C_R \,,\, S^C_R)$, we can write the terms above  as it follows,
\begin{equation}
{\cal L}=-\nu^C M_\nu \nu + {\mbox h.c.},
\label{ISSmatrix}
\end{equation}
where
\begin{equation}
M_{\nu}=
\begin{pmatrix}
0 & m^T_D & 0 \\
m_D & 0 & M_N^T\\
0 & M_N & \mu
\end{pmatrix},
\label{ISSmatrix}
\end{equation}
with $m_D$, $M_N$  and $\mu$ being $3\times 3$ mass matrices. Without loss of generality,  we consider $\mu$ diagonal and assume the following hierarchy $\mu << m_D << M_N$. We emphasize that after  block diagonalization of $M_\nu$ we obtain, in a first approximation, the following effective neutrino mass matrix for the standard neutrinos\cite{Hettmansperger:2011bt}:
\begin{equation}
m_\nu = m_D^T M_N^{-1}\mu (M_N^T)^{-1} m_D,
\label{inverseseesaw}
\end{equation}
while the heavy neutrinos obtain mass proportionally to $M_N$. We call this mechanism the inverse type I seesaw mechanism. There are two aspects that makes it profoundly distinct from the canonical case\cite{Minkowski:1977sc,Yanagida:1979as,GellMann:1980vs,Mohapatra:1979ia}, namely,  the double suppression by  $M_N$, which is an additional mass scale related to the six new right-handed neutral fermions, and the mass scale $\mu$, which is assumed to be very low\cite{Dias:2011sq}. The sub-eV active neutrino masses are then obtained by keeping $m_D$ at electroweak scale, $M_N$ at TeV scale and $\mu$ at keV scale. The new neutral fermions have their masses at TeV scale and their mixing with the standard neutrinos are modulated by the ratio 
$M_DM_N^{-1}$. The phenomenological appeal  of this  mechanisms is that it works at TeV scale and then  can be probed at the LHC\cite{Das:2012ze,Das:2014jxa,Das:2017zjc,Das:2017rsu,BhupalDev:2012zg}. One should notice that inverse seesaw  mechanisms seems more natural than the canonical one because $\mu \rightarrow 0$ enhances the symmetry of the model\cite{tHooft:1979rat}. For an implementation of this mechanism into the 3-3-1 model, see: \cite{Dias:2012xp}

%%%%%%%%%%%%%%%%%%%%%%%%%%%%%%%%%%%%%%%%%%%%

\subsection{Inverse type II seesaw mechanism}
%%%%%%%%%%%
Another approach for implementing  inverse seesaw mechanism into the standard model (SM) is by adding a triplet of scalars to the SM particle content\cite{Li:1985hy,Lusignoli:1990yk,deS.Pires:2005au,Freitas:2014fda},
\begin{equation}
\Delta\equiv \left(\begin{array}{cc}
\frac{\Delta^{+}}{\sqrt{2}} & \Delta^{++} \\ 
\Delta^{0} & \frac{-\Delta^{+}}{\sqrt{2}}
\end{array} \right)\,.
\end{equation}
The most simple gauge invariant potential composed by $\Delta$ and the standard scalar doublet  $\Phi = (\phi^{+}\,\,\,\, 
\phi^{0})^T$ involves the following terms:
\begin{eqnarray}
V(\Phi,\Delta) &=& -m_{H}^{2}\Phi^{\dagger}\Phi + \frac{\lambda}{4}(\Phi^{\dagger}\Phi)^{2} + M^{2}_{\Delta}Tr[(\Delta^{\dagger}\Delta)]+[\mu(\Phi^T i\sigma^{2}\Delta^{\dagger}\Phi)+H.c]\nonumber \\
&&+\lambda_{1}(\Phi^{\dagger}\Phi)Tr[(\Delta^{\dagger}\Delta)] +\lambda_{2}(Tr[(\Delta^{\dagger}\Delta)])^{2} +\lambda_{3}Tr[(\Delta^{\dagger}\Delta)^{2}] \nonumber \\
&&+ \lambda_{4}\Phi^{\dagger}\Delta^{\dagger}\Delta\Phi + \lambda_{5}\Phi^{\dagger}\Delta\Delta^{\dagger}\Phi\,.
\label{potential1}
\end{eqnarray}
Remember that $\Delta$ carries two units of lepton number. We draw attention to the trilinear term in the above potential. Perceive that this term  violates explicitly lepton number by two units.

In the SM perspective, when the neutral component of the doublet $\Phi$  develop a non zero vacuum expectation value (VEV) the electro-weak symmetry is broken and the mass of fermions arises as a result of the Higgs mechanism through the presence of Yukawa couplings of the fermion fields with the Higgs doublet.
By  similar way, the addition of the triplet $\Delta$, with its neutral component developing a nonzero VEV, opens the possibility of generating neutrino masses, as we are going to see further on.

In order to develop the scalar sector and obtain its spectrum, we re-parametrize the neutral components of $\Phi$ and  $\Delta$ in the usual way,
\begin{eqnarray}
 \phi^{0} , \Delta^{0} \rightarrow \frac{1}{\sqrt{2}}\left(  v_{\phi , \Delta} 
+R_{ _{\phi ,\Delta} } +iI_{_{\phi ,\Delta} }\right) \,.
\label{vacua} 
\end{eqnarray}
The VEV $v_\Delta$ modifies softly the $\rho$-parameter in the following way: $\rho=\frac{1+\frac{2v^2_\Delta}{v^2_\phi}}{1+\frac{4v^2_\Delta}{v^2_\phi}}$. The current value $\rho=1.0004^{+0.0003}_{-0.0004}$ \cite{Tanabashi:2018oca} implies the following upper bound $v_\Delta < 3$GeV. The regime of energy we are interested here is around eV for $v_\Delta$, which satisfy the upper bounds limits on $\rho$.

After the re-parametrization, the set of constraint equations,
\begin{eqnarray}
&&-m^{2}_{H}+\frac{1}{4}(v^{2}_{\phi}\lambda +2v_{\Delta}(v_{\phi}(\lambda_{1}+\lambda_{4})-2\sqrt{2}\mu))=0 ,\nonumber \\
&&M^{2}_{\Delta} v_{\Delta}+v^{3}_{\Delta}(\lambda_{2}+\lambda_{3})+\frac{v^{2}_{\phi}v_{\Delta}}{2}(\lambda_{1}+\lambda_{4}) -\frac{1}{\sqrt{2}}v^{2}_{\phi}\mu=0\,.
\label{constraints}
\end{eqnarray}
guarantee the potential above has a global  minimum.

Just like the type I seesaw mechanism, where the bare mass terms for the right-handed neutrinos are assumed to lie around keV scale, here we also  suppose that the parameter $\mu$ in  Eq.~(\ref{potential}) lies around keV scale, too.  As we will show,  this assumption implies a small VEV for $\Delta^0$.

The first constraint in Eq.~(\ref{constraints})  leads to $m^{2}_{H}\simeq\frac{1}{4}v^{2}_{\phi}\lambda$, while the second one provides,
\begin{equation}
 v_{\Delta}\simeq \frac{1}{\sqrt{2}}v_\phi M^{-1}_\Delta\mu v_\phi M^{-1}_\Delta .
 \label{ISSequation}
\end{equation}
The Eq.~(\ref{ISSequation}) is the main result of the inverse type II seesaw mechanism. As one can see, the parameter $v_\Delta$ gets suppressed due to the small energy scale associated to lepton number violation, $\mu$. In this manner, $v_\Delta$ at eV scale requires $M_\Delta$ around TeV  and $\mu$ around keV scale (while $v_\phi$ is the electroweak scale).

With  $\Delta$ and the standard leptonic doublet $L=(\nu \,\,,\,\, e)_L^T$ we have the following The Yukawa interactions,
\begin{equation}
\mathcal{L}_{Y}=Y_{ij}\bar{L}^{c}_{i} i \sigma_{2}\Delta L_{j} + H.c.
\label{Yukawacoupling}
\end{equation}
When $\Delta$ develops a VEV, $v_\Delta$, the Yukawa interactions provides the following expression to the neutrino mass,
\begin{equation}
m^\nu_{ij}=Y_{ij}v_{\Delta}=\frac{Y_{ij}}{\sqrt{2}}(v_\phi M^{-1}_\Delta)\mu (v_\phi M^{-1}_\Delta).
\label{neutrinomassII}
\end{equation}
 As interesting aspect emerging from this mechanism see that   Eq. ~(\ref{neutrinomassII}) recovers  Eq. (\ref{inverseseesaw}) with the advantage that now the  structure of the mass matrices becomes much more simple to handle. This neutrino mass generation mechanism is known as the inverse type II seesaw mechanism\cite{Li:1985hy,Lusignoli:1990yk,deS.Pires:2005au,Freitas:2014fda}. Its main signature are  singly and doubly charged scalars belonging to the triplet $\Delta$ whose mass values must  lie around the TeV scale and  therefore can be probed at the LHC\cite{Freitas:2014fda}.

\subsection{Spectrum of scalars}

From the scalar  potential in Eq.~(\ref{potential1}), and the constraint equations in Eq.~(\ref{constraints}), we obtain a $2 \times 2$  mass matrix  for the CP-even  neutral scalars.  Imposing the limit $v_{\phi}\gg \mu ,  v_\Delta$, the diagonalization of  this matrix provides the following eigenvalues,
\begin{equation}
\begin{split}
m^{2}_{h^{0}}&\simeq \frac{v^{2}_{\phi}\lambda}{4},\\
m^{2}_{H^{0}}&\simeq m^{2}_{h^{0}}+\left( \frac{1}{\sqrt{2}}\frac{\mu}{v_{\Delta}}\right)v^{2}_{\phi}.
\end{split}
\label{cpevenmass}
\end{equation}
with the respective eigenvectors,
\begin{equation}
\left(\begin{array}{c}
h^{0} \\ 
H^{0}
\end{array}\right)\simeq \left(\begin{array}{cc}
1 & \sqrt{\frac{v_{\Delta}}{v_{\phi}}} \\ 
- \sqrt{\frac{v_{\Delta}}{v_{\phi}}} & 1
\end{array}\right) \left(\begin{array}{c}
R_{\phi} \\ 
R_{\Delta}
\end{array}\right) ,
\end{equation}
where $h^0$ is the standard Higgs, while $H^0$ is a second Higgs that remains in the model. For $v_\Delta \approx 1$eV and $v_\phi \approx 10^2$GeV, we have $\frac{v_\Delta}{v_\phi}\approx 10^{-11}$.  In this case $h^0$ decouples from $H^0$. 

For the CP-odd neutral scalars the $2 \times 2$ mass matrix in the limit $v_{\phi}\gg \mu > v_{\Delta}$ gives us the following eigenvalues:
\begin{equation}
\begin{split}
m^{2}_{G^{0}}&=0, \\
m^{2}_{A^{0}}&\simeq \frac{1}{\sqrt{2}}v^{2}_{\phi}\frac{\mu}{v_{\Delta}},
\end{split}
\end{equation}
and their respective eigenvectors, 
\begin{equation}
\left(\begin{array}{c}
G^{0} \\ 
A^{0}
\end{array}\right)=\left(\begin{array}{cc}
\cos\beta & \sin\beta \\ 
-\sin\beta & \cos\beta
\end{array}\right)\left(\begin{array}{c}
I_{\phi} \\ 
I_{\Delta}
\end{array}\right),
\label{mixscalars}
\end{equation}
with
\begin{equation}\label{eq18}
\sin\beta =\frac{2v_{\Delta}}{\sqrt{v^{2}_{\phi}+2v^{2}_{\Delta}}}\,\,\,,\,\,\,
\cos\beta  =\frac{v_{\phi}}{\sqrt{v^{2}_{\phi}+2v^{2}_{\Delta}}}.
\end{equation}

Assuming $v_{\phi}\gg  v_\Delta$, we have $\sin\beta \rightarrow 0$ and $\cos\beta \rightarrow 1$ which implies that $G^0$ decouples from $A^0$. In this case we see that  $G^0$ is the Goldstone boson that will be absorbed by the SM neutral gauge boson $Z$,  and $A^0 $ is a massive CP-odd scalar that remain in the particle spectrum.

Regarding the singly charged scalars, their eigenvalues and eigenvectors are obtained from a $2\times 2$ mass matrix whose diagonalization provides, in the limit $v_{\phi}\gg v_{\Delta}, \mu$, the following eigenvalues:
\begin{equation}
\begin{split}
& m^{2}_{G^{+}}=0, \\
& m^{2}_{H^{+}}\simeq \frac{\sqrt{2}}{2}(\frac{\mu}{v_{\Delta}}-\lambda_{4})v^{2}_{\phi},
\end{split}
\label{hpmass}
\end{equation}
where $G^{+}$ is the Goldstone boson absorbed by the SM charged gauge bosons, $W^\pm$, while $H^\pm $ are massive scalars remaining in the spectrum. The mixing mass matrix for $\Delta^\pm$ and  $\phi^\pm$ is the same as in Eq.~(\ref{mixscalars}). Thus, in the limit $v_{\phi}\gg  v_\Delta$, the new singly charged scalars also decouples from the SM content.

In regard to the doubly charged scalars,  $\Delta^{\pm\pm}$,  we obtain the following expression for its mass in the limit $v_{\phi}\gg \mu , v_{\Delta} $,
\begin{equation}
m^{2}_{\Delta^{++}}\simeq \frac{\sqrt{2}}{2}(\frac{\mu}{v_{\Delta}}-\lambda_{4})v^{2}_{\phi}.
\label{hppmass}
\end{equation}
One should notice, besides  $\Delta$ belongs to the TeV energy scale, its particle content  decouples completely from the standard model scalar content. 

In summary, the inverse type II seesaw mechanism is a phenomenological viable seesaw mechanism whose signature are  scalars with mass at the TeV scale which concretely can be probed at the LHC\cite{Freitas:2014fda}.

\section{ The inverse type II seesaw mechanism and the 3-3-1 model with right-handed neutrinos}
\label{sec_III}
%%%%%
\subsection{Revisiting the model}
%%%%
The leptonic content of the model is arranged in a triplet and singlet of leptons  in the following form\cite{Singer:1980sw,Montero:1992jk,Foot:1994ym}
\begin{equation}
f_{aL}= \begin{pmatrix}
\nu_{a}     \\
\ell_{a}       \\
\nu^{c}_{a} \\
\end{pmatrix}_{L} \sim (1,3,-1/3), \quad e_{aR}\sim (1,1,-1),
\end{equation}
with $a=1,2,3$ representing the three SM generations of leptons.

In the Hadronic sector, the first generation comes in the triplet and the other two are in an anti-triplet, as a requirement to anomaly cancellation and are represented as follows,

\begin{eqnarray}
&&Q_{i_L} = \left (
\begin{array}{c}
d_{i} \\
-u_{i} \\
d^{\prime}_{i}
\end{array}
\right )_L\sim(3\,,\,\bar{3}\,,\,0)\,,u_{iR}\,\sim(3,1,2/3),\,\,\,\nonumber \\
&&\,\,d_{iR}\,\sim(3,1,-1/3)\,,\,\,\,\, d^{\prime}_{iR}\,\sim(3,1,-1/3),\nonumber \\
&&Q_{3L} = \left (
\begin{array}{c}
u_{3} \\
d_{3} \\
u^{\prime}_{3}
\end{array}
\right )_L\sim(3\,,\,3\,,\,1/3),u_{3R}\,\sim(3,1,2/3),\nonumber \\
&&\,\,d_{3R}\,\sim(3,1,-1/3)\,,\,u^{\prime}_{3R}\,\sim(3,1,2/3),
\label{quarks} 
\end{eqnarray}
where  the index $i=1,2$ is restricted to only two generations. The primed quarks are new heavy quarks with the usual $(+\frac{2}{3}, -\frac{1}{3})$ electric charges. 

The original scalar content of the 3-3-1 model carries three scalar triplets,
\begin{eqnarray}
\eta = \left (
\begin{array}{c}
\eta^0 \\
\eta^- \\
\eta^{\prime 0}
\end{array}
\right ),\,\rho = \left (
\begin{array}{c}
\rho^+ \\
\rho^0 \\
\rho^{\prime +}
\end{array}
\right ),\,
\chi = \left (
\begin{array}{c}
\chi^0 \\
\chi^{-} \\
\chi^{\prime 0}
\end{array}
\right ),
\label{scalarcont} 
\end{eqnarray}
with $\eta$ and $\chi$ transforming as $(1\,,\,3\,,\,-1/3)$
and $\rho$ as $(1\,,\,3\,,\,2/3)$.

The Yukawa lagrangian of the model is described by the following terms:
\begin{eqnarray}
&-&{\cal L}^Y =f_{ij} \bar Q_{i_L}\chi^* d^{\prime}_{j_R} +f_{33} \bar Q_{3_L}\chi u^{\prime}_{3_R} + g_{ia}\bar Q_{i_L}\eta^* d_{a_R} \nonumber \\
&&+h_{3a} \bar Q_{3_L}\eta u_{a_R} +g_{3a}\bar Q_{3_L}\rho d_{a_R}+h_{ia}\bar Q_{i_L}\rho^* u_{a_R}+ G_{l}\bar f_{l_L} \rho e_{l_R} + \mbox{H.c},
\label{yukawa}
\end{eqnarray}
where $a=1,2,3$. For the sake of simplicity, we considered the charged leptons in a diagonal basis.

The 3-3-1 model recovers the standard gauge bosons, as a consequence of the model we have the addition of five more vector bosons called  $V^{\pm}$, $U^0$, $U^{0 \dagger}$ and $Z^{\prime}$.

The most general gauge invariant potential is given by the following terms:
\begin{eqnarray} 
V^{\prime}(\eta,\rho,\chi)&=&\mu_\chi^2 \chi^2 +\mu_\eta^2\eta^2
+\mu_\rho^2\rho^2+\lambda_1\chi^4 +\lambda_2\eta^4
+\lambda_3\rho^4+ \nonumber \\
&&\lambda_4(\chi^{\dagger}\chi)(\eta^{\dagger}\eta)
+\lambda_5(\chi^{\dagger}\chi)(\rho^{\dagger}\rho)+\lambda_6
(\eta^{\dagger}\eta)(\rho^{\dagger}\rho)+ \nonumber \\
&&\lambda_7(\chi^{\dagger}\eta)(\eta^{\dagger}\chi)
+\lambda_8(\chi^{\dagger}\rho)(\rho^{\dagger}\chi)+\lambda_9
(\eta^{\dagger}\rho)(\rho^{\dagger}\eta) \nonumber \\
&&-\frac{f}{\sqrt{2}}\epsilon^{ijk}\eta_i \rho_j \chi_k +\mbox{H.c}.
\label{potential}
\end{eqnarray}

The simplest scenario is when the VEV structure  of the 3-3-1 model comes in a diagonal form, i.e. only $\eta^0$, $\rho^0$ and $\chi^{\prime 0}$ develop VEV. In this case, considering the expansions around the VEV:
\begin{eqnarray}
 \eta^0 , \rho^0 , \chi^{\prime 0} \rightarrow  \frac{1}{\sqrt{2}} (v_{\eta ,\rho ,\chi^{\prime}} 
+R_{ \eta ,\rho ,\chi^{\prime}} +iI_{\eta ,\rho ,\chi^{\prime}}),
\label{vacua} 
\end{eqnarray}
Replacing E.q.~(\ref{vacua}) in the potential (\ref{potential}), we get the set of constrains:
\begin{eqnarray}
 &&\mu^2_\chi +\lambda_1 v^2_{\chi^{\prime}} +
\frac{\lambda_4}{2}v^2_\eta  +
\frac{\lambda_5}{2}v^2_\rho-\frac{f}{2}\frac{v_\eta v_\rho}
{ v_{\chi^{\prime}}}=0,\nonumber \\
&&\mu^2_\eta +\lambda_2 v^2_\eta +
\frac{\lambda_4}{2} v^2_{\chi^{\prime}}
 +\frac{\lambda_6}{2}v^2_\rho -\frac{f}{2}\frac{v_{\chi^{\prime}} v_\rho}
{ v_\eta} =0,
\nonumber \\
&&
\mu^2_\rho +\lambda_3 v^2_\rho + \frac{\lambda_5}{2}
v^2_{\chi^{\prime}} +\frac{\lambda_6}{2}
v^2_\eta-\frac{f}{2}\frac{v_\eta v_{\chi^{\prime}}}{v_\rho} =0.
\label{mincond} 
\end{eqnarray}

The 3-3-1 model recovers all the predictions of the standard model and provides a set of new physics predictions which can be probed at the LHC. However, the model, in its original form, does not address mass to the  neutrino content. One interesting way of including mass terms to the neutrinos is by implementing the inverse type II seesaw mechanism into its framework. 

\subsection{Implementing the type II seesaw mechanism into the  3-3-1 model}

The implementation of the mechanism into the 3-3-1 model requires the addition of a sextet of scalars\cite{Long:2016lmj} to the original scalar content of the 3-3-1 model,
\begin{equation}
\label{SextetoEscalar}
S=\frac{1}{\sqrt{2}}\left(\begin{array}{ccc}
\sqrt{2}\, \Delta^{0} & \Delta^{-} & \Phi^{0} \\
\newline \\
\Delta^{-} & \sqrt{2}\, \Delta^{--} & \Phi^{-} \\
\newline \\
\Phi^{0} & \Phi^{-} & \sqrt{2}\, \sigma^{0} \end{array}\right) \sim (1,6,-2/3).
\end{equation}
With $S$ and the triplets $f$ we form the following Yukawa interaction,
\begin{equation}
\label{ysexteto}
{\cal L}^{\nu}_{Y}= G_{ab} \overline{f_{aL}}\, S\,(f_{bL})^c + H.C.
\end{equation}

%(Expanding it we get,
%\begin{eqnarray}
%{\cal L}^{\nu}_{Y}& = & \frac{G_{ab}}{\sqrt{2}} \left[ \overline{\nu_{aL}}\,\sqrt{2}\,\Delta^0 \,(\nu_{bL})^c + \overline{\ell_{aL}}\,\Delta^- \,(\nu_{bL})^{c} + \overline{(\nu^c_a)}_L\,\Phi^0 \,(\nu_{bL})^c \right] \nonumber \\
%&+& \frac{G_{ab}}{\sqrt{2}}\left[ \overline{\nu_{aL}}\,\Delta^- \,(\ell_{bL})^c + \overline{\ell_{aL}}\,\sqrt{2}\,\Delta^{--} \,(\ell_{bL})^{c} + \overline{(\nu^c_a)}_L\,\Phi^- \,(\ell_{bL})^c \right]  \nonumber \\
%&+& \frac{G_{ab}}{\sqrt{2}}\left[ \overline{\nu_{aL}}\,\Phi^0 \,\nu_{bR} + \overline{\ell_{aL}}\,\Phi^- \,\nu_{bR} + \overline{(\nu^c_a)}_L\,\sqrt{2}\,\sigma^0 \,\nu_{bR} \right] + H.C.\, 
%\label{yul}
%\end{eqnarray})

Observe that  when we assume that only  $\Delta^0$ and $\sigma^0$ develop VEV  different from zero, this Yukawa interaction provides the following mass terms for the  left-handed and right-handed neutrinos,
\begin{eqnarray}
{\cal L}^{\nu}_{Y} = \frac{\sqrt{2}\,G_{ab}v_\Delta}{2}\,\, \overline{\nu_{aL}}\, \,(\nu_{bL})^c + \frac{\sqrt{2}\,G_{ab}v_\sigma}{2}\,\, \overline{(\nu_{aR})^c}\,\,\nu_{bR} + H.C.\, .
\label{neutrino_mass}
\end{eqnarray}
In this point it turns important to check if the minimal condition constraints over the VEVs allow  such choice. 

In order to have the simplest gauge invariant potential that violate explicitly lepton number, which is necessary for we have the seesaw mechanism, we resort to a $Z_3$ discrete symmetry  with the fields of interest transforming as $(S, f)\rightarrow w(S,f)$ and $(\eta,\chi) \rightarrow w^{-1}(\eta,\chi)$ where $w=e^{2 \pi i/3}$.  In this case, the new potential involves the following terms:
\begin{eqnarray}
\label{Potencial}
V&=&  \mu^{2}_{\chi}\chi^{\dagger}\chi + \mu^{2}_{\eta}\eta^{\dagger}\eta + \mu^{2}_{\rho}\rho^{\dagger}\rho + \lambda_{1}(\chi^{\dagger}\chi)^{2} + \lambda_{2}(\eta^{\dagger}\eta)^{2} + \lambda_{3}(\rho^{\dagger}\rho)^{2}     \nonumber  \\
&+& \lambda_{4}(\chi^{\dagger}\chi)(\eta^{\dagger}\eta) + \lambda_{5}(\chi^{\dagger}\chi)(\rho^{\dagger}\rho) + \lambda_{6}(\eta^{\dagger}\eta)(\rho^{\dagger}\rho) + \lambda_{7}(\chi^{\dagger}\eta)(\eta^{\dagger}\chi)         \nonumber  \\
&+& \lambda_{8}(\chi^{\dagger}\rho)(\rho^{\dagger}\chi) +\lambda_{9}(\eta^{\dagger}\rho)(\rho^{\dagger}\eta)   \nonumber \\
&+& \mu^{2}_{S}Tr\left(S^{\dagger}S \right) + \lambda_{10}Tr\left[(S^{\dagger}S \right)^2] + \lambda_{11}\left[Tr(S^{\dagger}S) \right]^{2} \nonumber \\
&+& \left[\lambda_{12}(\eta^{\dagger}\eta) + \lambda_{13}(\rho^{\dagger}\rho) + \lambda_{14}(\chi^{\dagger}\chi) \right] Tr\left(S^{\dagger}S \right)         \nonumber \\
&+& \lambda_{15}\left( \chi^{\dagger}S\right) \left( S^\dagger \chi \right) + \lambda_{16}\left(\eta^{\dagger}S\right) \left(S^\dagger\eta\right) + \lambda_{17}\left(\rho^{\dagger}S\right) \left(S^\dagger\rho\right)           \nonumber \\
&-&M_1 \eta^T S^\dagger \eta - M_2 \chi^T S^\dagger \chi + H.C.  
\end{eqnarray}

Assuming that only $ \Delta^0 $  and $\sigma^0 $  develop VEVs, and expanding them around their VEVs in the usual way,
\begin{eqnarray}
 \Delta^0 , \sigma^0 \rightarrow  \frac{1}{\sqrt{2}} (v_{\Delta ,\sigma } 
+R_{ \Delta ,\sigma} +iI_{\Delta ,\sigma }), 
\label{vacuaII} 
\end{eqnarray}
then the potential above provides the following set of  constraint equations,
\begin{eqnarray}
\label{minimum_pot}
&&\mu^2_{\chi\prime} + \lambda_1 v_{\chi\prime}^2 + \frac{\lambda_4}{2}v_{\eta}^2 + \frac{\lambda_5}{2}v_{\rho}^2 + \frac{\lambda_{14}}{4}\left(v_{\sigma}^2 + v_{\Delta }^2 \right) + \frac{\lambda _{15}}{4}v_{\sigma}^2 - M_2 v_{\sigma} = 0,  \nonumber    \\
&&\mu^2_{\eta} + \lambda_2 v_{\eta}^2 + \frac{\lambda_4}{2} v_{\chi\prime}^2 + \frac{\lambda_6}{2} v_{\rho}^2 + \frac{\lambda_{12}}{4} \left(v_{\sigma}^2 + v_{\Delta}^2 \right) + \frac{\lambda _{16}}{4} v_{\Delta}^2 - M_1 v_{\Delta} = 0, \nonumber  \\
&&\mu^2_{\rho} + \lambda_3 v_{\rho}^2 +\frac{\lambda_5}{2} v_{\chi\prime}^2 + \frac{\lambda_6}{2} v_{\eta}^2 + \frac{\lambda_{13}}{4} \left(v_{\sigma}^2 + v_{\Delta}^2 \right) = 0,\\
&&\mu^2_S + \frac{\lambda_{10}}{2} v_{\Delta }^2 +\frac{\lambda_{11}}{2} \left( v_{\Delta}^2 + v_{\sigma}^2 \right) + \frac{\lambda_{12}}{2} v_{\eta}^2 +\frac{\lambda_{13}}{2} v_{\rho}^2 + \frac{\lambda_{14}}{2} v_{\chi\prime}^2 + \frac{\lambda_{16}}{2} v_{\eta}^2 - M_1 \frac{v_{\eta}^2}{v_\Delta} = 0,  \nonumber \\
&&\mu^2_S + \frac{\lambda_{10}}{2} v_{\sigma}^2 + \frac{\lambda_{11}}{2} \left(v_{\sigma}^2 + v_{\Delta}^2 \right) + \frac{\lambda_{12}}{2} v_{\eta}^2 + \frac{\lambda_{13}}{2} v_{\rho}^2 + \frac{\lambda_{14}}{2} v_{\chi\prime}^2 + \frac{\lambda_{15}}{2} v_{\chi\prime}^2  - M_2 \frac{v_{\chi\prime}^2}{v_\sigma} = 0, \nonumber
\end{eqnarray}
 which guarantee that the potential develops  a global minimum. 

As the sextet $S$ is an extension of the 3-3-1 model, it is natural to expect that its content be heavier than the original 3-3-1 scalar  triplets. In view of this, the parameter  $\mu_S$ in the last two relations in Eq. (\ref{minimum_pot}) must dominate over the other ones, except the last ones ($M_1 \frac{v_{\eta}^2}{v_\Delta}\,\, \mbox{and}\,\, M_2 \frac{v_{\chi\prime}^2}{v_\sigma}$). Consequently, the last two relations provide the following expressions for the VEVs $v_\sigma$ and $v_\Delta$:
\begin{equation}
v_\Delta \simeq  M_1\,\frac{v^2_\eta}{\mu^2_s}, \quad v_\sigma \simeq  M_2\,\frac{v^2_{\chi^\prime}}{\mu^2_s}.
\label{vevrel}
\end{equation}
This is the inverse type II seesaw mechanism where  tiny VEVs, $v_\Delta$ and $v_\sigma$, is an implication of the smallness of the parameters $M_1$ and $M_2$. Observe that  $v_{\chi^\prime} > v_\eta$ implies $v_\sigma > v_\Delta$ which generates a hierarchy among: $v_\sigma> v_\Delta$.

Replacing the above expressions for the VEVs $v_\Delta$ and $v_\sigma$ in Eq. (\ref{neutrino_mass}), we obtain the following expressions for the neutrinos masses, 
\begin{equation}
\label{mass_neu}
m_{\nu_L} = \frac{\,G}{\sqrt{2}}\,(v_\eta \mu_S^{-1})M(v_\eta \mu^{-1}_S), \quad m_{\nu_R} = \frac{\,G}{\sqrt{2}}\,(v_{\chi^\prime}\mu_S^{-1})M(v_{\chi^\prime}\mu^{-1}_S).
\end{equation}
 These expressions  are remarkably similar to the one in Eq. (\ref{neutrinomassII}). Observe, also, that the left-handed and right-handed neutrino masses share the same Yukawa coupling, $G$. Consequently, fixing the masses of the left-handed neutrinos automatically we have the masses of the  right-handed neutrinos.

 The energy parameters $M_1$ and $M_2$ are associated to the explicit violation of the  lepton number. In this way, observe  that the  potential get more symmetric when $M_1$ and $M_2$ go to zero. In other words,  the smaller $M_1$ and $M_2$ are, more symmetric the potential is. For simplification reasons, let us assume $M_1=M_2=M$. Inverse seesaw mechanisms require lepton number be explicitly violate at low energy scale. Here is not different. For instance, for $$\mu_S \sim 10^3\,\text{GeV}, \quad v_{\chi\prime}\sim 10^4\,\text{GeV},\quad M= 10^{-2}\text{keV}\quad  \text{and}\quad v_{\eta}\sim 10^2\,\text{GeV},$$ Eq. (\ref{mass_neu}) give us
\begin{equation}
m_{\nu_L} = \frac{G}{\sqrt{2}}10^{-10}\,\text{GeV}, \quad m_{\nu_R} = \frac{G}{\sqrt{2}} 10^{-6}\,\text{GeV}.
\end{equation}
 
The current set of experimental data in neutrino physics is not enough to fix all the Yukawa coupling, $G$'s. Despite that, as illustrative example, we chose  $G_{11}= -0.0023, G_{12} = 0.0012, G_{13}= 0.0049, G_{22}= -0.026, G_{23}= -0.0170, G_{33}= -0.0189$ as primary set of benchmark points. With this set of values for the Yukawa couplings, the diagonalization of $m_{\nu_L}$ in Eq. (\ref{mass_neu}) produces the following  masses for the physical standard neutrinos
(in the normal hierarchy):
\begin{eqnarray}
m_{N_1}&\approx& 0,  \nonumber\\
m_{N_2}&=& 8.61 \times 10^{-3}\,\text{eV},  \\
m_{N_3}&=& 5.0\times 10^{-2}\,\text{eV}.  \nonumber 
\end{eqnarray}
These predictions provide the following mass differences:
\begin{eqnarray}
(\Delta m_{21})^{2} &=& 7.50 \times 10^{-5}\,\text{eV}^{2} \quad  (\textit{solar})\nonumber\\
(\Delta m_{31})^{2} &=& 2.524 \times 10^{-3}\,\text{eV}^{2} \quad  (\textit{atmospheric}),
\end{eqnarray}
which explain solar and atmospheric neutrino oscillation experimental results \cite{Tanabashi:2018oca}.
 
 Because $m_{\nu_L}$ and $m_{\nu_R}$, both, share the same set of Yukawa couplings, then the mechanism predicts right-handed neutrinos with the following masses,
\begin{eqnarray}
m_{S_1}&\approx & 2.65\times 10^{-3} \,\, \,\text{eV}, \nonumber\\
m_{S_2}&\approx & 21.5 \,\,  \text{eV}, \\
m_{S_3}&\approx & 125\,\,  \text{eV}. \nonumber
\end{eqnarray}

Right-handed neutrinos are naturally heavy particle because they are singlet by the standard symmetry. Here we are obtaining a interesting result that is light-right-handed neutrinos. This is a astonishing result. 

In what concern  the relations between flavor and mass eigenstates,  we have,
\begin{eqnarray}
\nu_{L} = U_{PMNS}\,N_{L},\quad (\nu_R)^{c} = U_{R}\,(S_R)^{c}, \end{eqnarray}
where $N_L = (N_1 , N_2 , N_3 )^T$ and $S_R = (S_1 , S_2 , S_3 )^T$.

For the set of Yukawa couplings choose above, the respective mixing matrices for the left and right-handed  neutrinos  are given by:
\begin{equation}
U_{PMNS}=\left(\begin{array}{ccc}
  0.830   &  0.540   &   -0.120       \\
\newline \\
  -0.250  &  0.590   &   0.720        \\
\newline \\
  0.440   &  -0.600  &   0.690        \\
\end{array}\right),\quad
U_{R}=\left(\begin{array}{ccc}
 0.854     &     0.509    &    0.107  \\
\newline \\
 -0.257    &     0.592    &     -0.764\\
\newline \\
 0.453     &    -0.625    &     -0.636\\
\end{array}\right).
\end{equation}

Thus, we succeeded to implement the inverse type II seesaw mechanism into the 3-3-1 model with right-handed neutrinos. As nice result the mechanism predicts tiny masses for both left-handed and right-handed neutrinos.

\subsection{The Spectrum of scalars }

After the implementation of the mechanism, the next point is to present and develop its signature.  In this perspective, it is a mandatory step to develop the potential in Eq.~(\ref{Potencial}) considering the minimum condition equations in Eq. (\ref{minimum_pot}). Basically we check the mixture among the scalars that belong to the sextet with the original content of scalars of the model. This just requires we present the mass matrices in their correct basis. 

Let us begin with the CP-even neutral scalars. As basis we take $R=(R_{\chi\prime},R_\eta,R_\rho,R_\Delta,R_\sigma,R_\Phi,R_\chi,R_{\eta\prime})$. In this case we obtained the following mass matrices for the CP-even neutral scalars:
%%%
\begin{equation}
\label{}
M_R \simeq \left(\begin{array}{ccc}
M_{\chi\prime,\eta,\rho} & M^{\prime} & 0 \\
\newline \\
M^{\prime} & M_{\Delta,\sigma,\Phi} & 0   \\
\newline \\
0 & 0 & M_{\chi,\eta\prime}
\end{array}\right),
\end{equation}
where,
\begin{eqnarray}
\label{Mix_chi_eta_rho}
&&M_{\chi\prime,\eta,\rho} = \left(\begin{array}{lll}
\lambda_1 v_{\chi\prime}^2 \,\, & \lambda_4 v_{\chi\prime} v_{\eta} \,\, & \lambda_5 v_{\chi\prime} v_{\rho}  \\
\newline \\
\lambda_4 v_{\chi\prime} v_{\eta} \,\, & \lambda_2 v_{\eta}^2 \,\, & \lambda_6 v_{\eta}v_\rho  \\
\newline \\
\lambda_5 v_{\chi\prime} v_{\rho} \,\, & \lambda_6 v_{\eta}v_\rho \,\, & \lambda _3 v_{\rho }^2  
\end{array}\right), \,\,\,\,
M_{\Delta,\sigma,\Phi} = \left(\begin{array}{lll}
M^2_{R_\Delta} & \frac{\lambda_{11}}{2} v_{\Delta} v_{\sigma} & 0 \\
\newline \\
\frac{\lambda_{11}}{2} v_{\Delta} v_{\sigma} \,\, & M^2_{R_\sigma} \,\, & 0 \\
\newline \\
0 & 0 & M^2_{R_\phi} \\
\end{array}\right),\nonumber \\
&& M_{\chi,\eta\prime} = \left(\begin{array}{ll}
M^2_{R_\chi} & 0 \\
\newline \\
0 & M^2_{R_{\eta\prime}}
\end{array}\right)\,\,\, \mbox{and}\,\,\,\,\, M^{\prime} = \left(\begin{array}{lll}
 \frac{\lambda_{14}}{2} v_{\chi\prime} v_{\Delta} & \Lambda_1 \,v_{\chi\prime} & 0 \\
\newline \\
 \Lambda_2 \,v_{\eta} & \frac{\lambda_{12}}{2} v_{\eta} v_{\sigma}  & 0 \\
\newline \\
 \frac{\lambda_{13}}{2} v_{\rho} v_{\Delta} & \frac{\lambda_{13}}{2} v_{\rho} v_{\sigma}   & 0 \\
\end{array}\right),
\end{eqnarray}
with
\begin{eqnarray}
&&M^2_{R_\Delta} = \frac{ \lambda _{10}}{4} v_\Delta^2 + M \frac{v_{\eta}^2}{4v_\Delta},\\
&&M^2_{R_\sigma} = \frac{\lambda_{10}}{4}v_\sigma^2 + M \frac{v_{\chi\prime}^2}{4v_\sigma},\\
&&M^2_{R_\phi} = \frac{\lambda_{10} }{8} \left( v_\sigma^2 - v_\Delta^2 \right) - \frac{\lambda _{16}}{8}v_\eta^2 +  \frac{\lambda _{15} }{8}v_{\chi\prime}^2 + M \frac{v_{\eta}^2}{2v_\Delta},\\
&&M^2_{R_\chi} = M^2_{R_\chi} = \frac{1}{2} \left[ M (v_\sigma - v_\Delta) + \frac{\lambda_{15}}{4}(v_\Delta^2 - v_\sigma^2) +\frac{\lambda_7}{2} v_\eta^2 \right],\\
&&M^2_{R_{\eta\prime}} = \frac{1}{2} \left[ M ( v_\Delta - v_\sigma ) + \frac{\lambda _{16}}{4}(v_\sigma^2 - v_\Delta^2) + \frac{\lambda _7}{2} v_{\chi\prime}^2\right],
\end{eqnarray}
and
$$\Lambda_1 = \frac{1}{2}\left( \lambda_{14} + \lambda_{15}\right) v_{\sigma} - M, \quad \Lambda_2 = \frac{1}{2}\left( \lambda_{12} + \lambda_{16}\right) v_{\Delta} - M, $$ $$ \Lambda_3 = \frac{\lambda_{15}}{4}\left( v_{\Delta} + v_{\sigma}\right) - M, \quad \Lambda_4 = \frac{\lambda_{16}}{4}\left( v_{\Delta} + v_{\sigma}\right) - M. $$

One should notice that in our case  $v_\sigma,\,\, v_\Delta$ and  $M$ are much smaller than $v_\eta,\,\,\, v_\rho$ and $v_{\chi^{\prime}}$. Then we have: 
\begin{equation}
\label{}
M_R \simeq \left(\begin{array}{ccc}
M_{\chi\prime,\eta,\rho} & 0 & 0 \\
\newline \\
0 & M_{\Delta,\sigma,\Phi} & 0   \\
\newline \\
0 & 0 & M_{\chi,\eta\prime}
\end{array}\right). 
\end{equation}
This means that  $R_\Delta, R_\sigma, R_\Phi$ decouples from the original scalar content of the 3-3-1 model. This is a remarkable result once will facilitate the search for the signature of the mechanism in the form of doubly charged scalars. We do not show here but this behavior is kept with the CP-odd neutral scalar content.

For the singly charged scalar fields, considering the basis ($\eta^{-},\rho^+, \rho^{\prime +},\chi^-,\phi^{-},\Delta^{-})$, so we obtain:

\begin{equation}
\label{base_flavor_scalar}
M^{\pm} = \left(\begin{array}{cccccc}
M^2_{\eta^\pm}  & \frac{1}{2}\lambda_{9}v_{\eta}v_{\rho}  & 0 & 0 & 0 & \Lambda_{6}v_{\eta} \\
\newline \\
\frac{1}{2}\lambda_{9}v_{\eta}v_{\rho} & M^2_{\rho^\pm} & 0 & 0 & 0 & \frac{1}{2}\lambda_{17}v_{\rho}v_{\Delta} \\
\newline \\
0 & 0 & M^2_{\rho^{\prime\pm}} & \frac{1}{2}\lambda_{8}v_{\chi\prime}v_{\rho} & \frac{1}{4}\lambda_{17}v_{\rho}v_{\sigma} & 0 \\ 
\newline \\
0 & 0 & \frac{1}{2}\lambda_{8}v_{\chi\prime}v_{\rho} & M^2_{\chi^\pm} & \Lambda_{5}v_{\chi\prime} & 0 \\
\newline \\
0 & 0 & \frac{1}{4}\lambda_{17}v_{\rho}v_{\sigma} & \Lambda_{5}v_{\chi\prime} & M^2_{\phi^\pm} & 0 \\
\newline \\
\Lambda_{6}v_{\eta} & \frac{1}{2}\lambda_{17}v_{\rho}v_{\Delta} & 0 & 0 & 0 & M^2_{\Delta^\pm} \\
\end{array}\right),
\end{equation}
with 
\begin{eqnarray}
&&M^2_{\eta^\pm} = \frac{\lambda _9 }{2}v_\rho^2 - \frac{\lambda _{16}}{4} v_{\Delta}^2 + M v_{\Delta}
,\\
&&M^2_{\rho^\pm} = \frac{\lambda _{17}}{4} v_\Delta^2 + \frac{\lambda _9 }{2}v_\eta^2,\\
&&M^2_{\chi^\pm} = \frac{\lambda _8}{2}v_\rho^2 - \frac{\lambda _{15}}{4}v_{\sigma}^2 + M v_{\sigma},\\
&&M^2_{\rho^{\prime\pm}} = \frac{\lambda _{17}}{4}v_\sigma^2+\frac{\lambda _8 }{2}v_{\chi\prime}^2,\\
&&M^2_{\phi^\pm} = \frac{\lambda _{17}}{4}v_{\rho}^2-\frac{\lambda _{10} }{4}v_{\sigma}^2 -\frac{\lambda _{15}}{4}v_{\chi\prime}^2 + M \frac{v_{\chi\prime}^2}{v_\sigma},\\
&&M^2_{\Delta^\pm} = \frac{\lambda _{17}}{4} v_{\rho}^2 + M \frac{v_{\eta}^2}{v_\Delta},
\end{eqnarray}
and
\begin{eqnarray}
\Lambda_5 = \frac{1}{4}\lambda_{15}(v_\sigma - 4M),\quad
\Lambda_6 = \frac{1}{4}\lambda_{16}(v_{\Delta} - 4M).
\end{eqnarray}
The elements $M^{\pm}_{1\times 6}$, $M^{\pm}_{2\times 6}$, $M^{\pm}_{3\times 5}$ and $M^{\pm}_{4\times 5}$ are proportional to $M$, $v_\Delta$ and $v_\sigma$  which means that the singly charged scalars from the sextet decouples from the singly charged scalars of the original 3-3-1 triplets. 

Finally, with help of the fourth expression in \eqref{minimum_pot}, we can write
\begin{eqnarray*}
m^2_{\Delta^{++}} = \frac{1}{2} \left( \frac{\lambda_{17}}{2} v_{\rho }^2 -\frac{\lambda_{16}}{2} v_{\eta}^2  - \frac{\lambda_{10}}{2} v_{\Delta}^2 + M \frac{v_{\eta}^2}{v_\Delta}  \right).
\end{eqnarray*}

In summary, in the regime of energy  of validity of  the inverse type II seesaw mechanism, the particle content of the sextet of scalars decouples from the original scalar content of the 3-3-1 model. 

This analysis allows we conclude that the doubly charged scalar  turn out to be  the best candidate for the signature of the model. With this in mind, in the next section we present the prospects for detecting such scalar at the LHC. For this we consider a production of two doubly charged scalar in the resonance of $Z^{\prime}$ channel, followed by its dominant decay into pairs of leptons. 
%%%%%%%%%%%%%%%%%%%%%%%%
\section{Probing the signature of the mechanism at the LHC}
\label{sec_IV}
The doubly charged scalars $\Delta^{\pm \pm}$  can be probed at the LHC by producing and detecting their final decay states. In this section we restrict our investigation to the process $pp \rightarrow Z\,,\,\gamma\,,\,Z^{\prime}\rightarrow \Delta^{++}\Delta^{--}$. As final product, we consider  pairs of leptons and anti-leptons. Moreover, in order to enhance the production of $\Delta^{\pm\pm}$ we estimate the cross section of the process in the resonance of $Z^{\prime}$. This is justified due to the fact that  the detection of $\Delta^{++}$ in the form of leptons as final product  involves tiny Yukawa couplings $G_{ij}$. Our analysis is done for the specific setting of Yukawa couplings considered in the example presented before: $G_{11}= -0.0023, G_{12} = 0.0012, G_{13}= 0.0049, G_{22}= -0.026, G_{23}= -0.0170, G_{33}= -0.0189$. As a first approach, we present the behavior of the cross sections involved in the process by scanning $m_\Delta$ and $m_{Z^{\prime}}$. We also discuss the relevance of  $Z^{\prime}$ channel in relation to the standard $\gamma$ and $Z$ channels.  We estimate the profile likelihood for the identification of the doubly charged scalar and the $Z^{\prime}$ over the SM four leptons background. 
In order to simulate the necessary data of our events, we make use of the package FeynRules \cite{Alloul:2013bka} which provide as an output UFO\cite{Degrande:2011ua} which is used by Madgraph5 \cite{Degrande:2011ua} to generate the events. 
%%%%%%%%%%%%%
\begin{figure}[ht]
\centering
\includegraphics[scale=0.22]{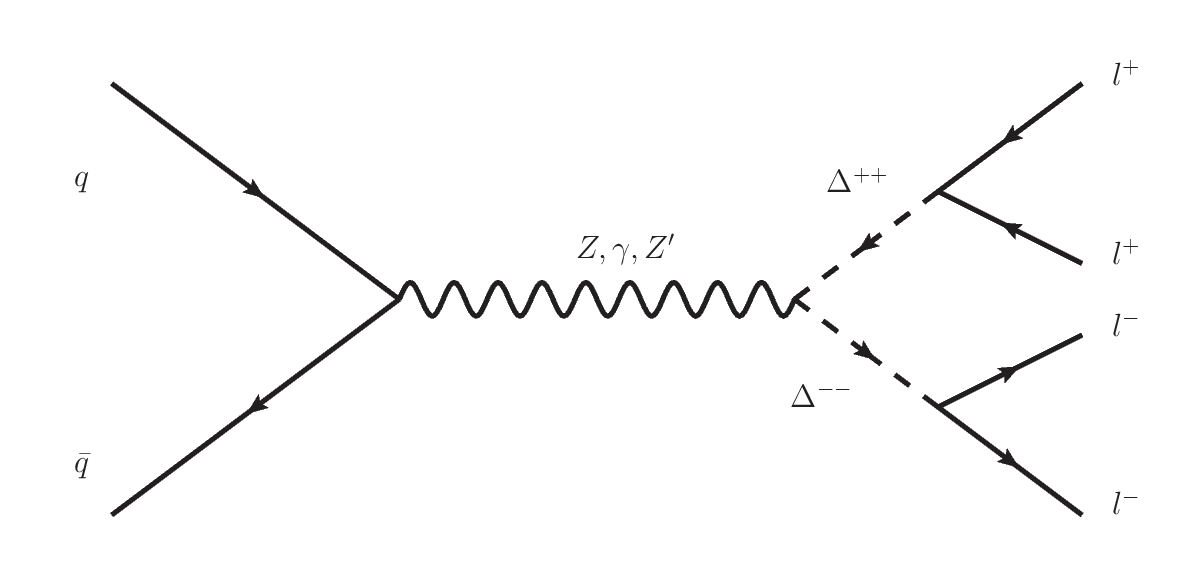}
\caption{Feynman diagram for the process $pp \rightarrow Z\,,\,\gamma\,,\,Z^{\prime}\rightarrow \Delta^{++}\Delta^{--}$ with the subsequent $\Delta^{\pm\pm}$ decays into leptons}
\label{process_pp_4l}
\end{figure}
%%%%%%%%%%%

 For a small $v_\Delta$, which is our case,  we have that $\Delta^{++}$ will  decay dominantly in pair of leptons\cite{Perez:2008ha}. For the set of Yukawa couplings of the example above, the branching ratio of the decay of $\Delta^{--}$ in pair of electrons  is very small, while the decay into pairs of taus is dominant. However, due to the low tau tag efficiency and the fact that taus prefer to decay into hadrons all this  make this channel a poor choice to properly reconstruct $\Delta^{\pm\pm}$ and $Z^{\prime}$. In view of this, the best choice seems to be to consider the decay of $\Delta^{--}$ into pairs of muons. This is a reasonable choice since the model predicts the following branching ratio for this channel:
\begin{equation}
BR(\Delta^{\pm\pm} \rightarrow \mu^\pm \mu^\pm)\simeq 0.41,
\label{mainBR}
\end{equation}
for $m_{\Delta^{++}}=700$\,GeV. 

The contribution from the Higgs boson to the process we are investigating is too small, and we can safely neglect it. Considering the  production of $\Delta^{\pm\pm}$ through the neutral SM gauge bosons and $Z^{\prime}$ for the case of  $m_{Z^\prime}=4$\,TeV we have the following cross sections for the LHC running with center of mass energy of 14\,TeV, 28\,TeV and 100\,TeV, respectively:
\begin{eqnarray}
&&\sigma (p\,p \rightarrow Z^*, \gamma^* ,Z^{\prime} \rightarrow \Delta^{++}\,\Delta^{--}) \sim 0.06\,\text{fb}, \\ \nonumber 
&&\sigma (p\,p \rightarrow Z^* , \gamma^* , Z^{\prime} \rightarrow \Delta^{++}\,\Delta^{--}) \sim 0.45\,\text{fb}, \\ \nonumber
&&\sigma (p\,p \rightarrow Z^* , \gamma^* ,Z^{\prime} \rightarrow \Delta^{++}\,\Delta^{--}) \sim 5.5\,\text{fb},
\end{eqnarray}
As one could expected, the cross section for 14 \,TeV is small, compared to 28 \,TeV and 100 \,TeV, and the detection of $Z^{\prime}$ and $\Delta^{\pm\pm}$ would require great amount of luminosity. However, the cross section for 28\,TeV and 100\,TeV are encouraging and justify go further.  

\begin{figure}[ht]
\centering
\includegraphics[scale=0.22]{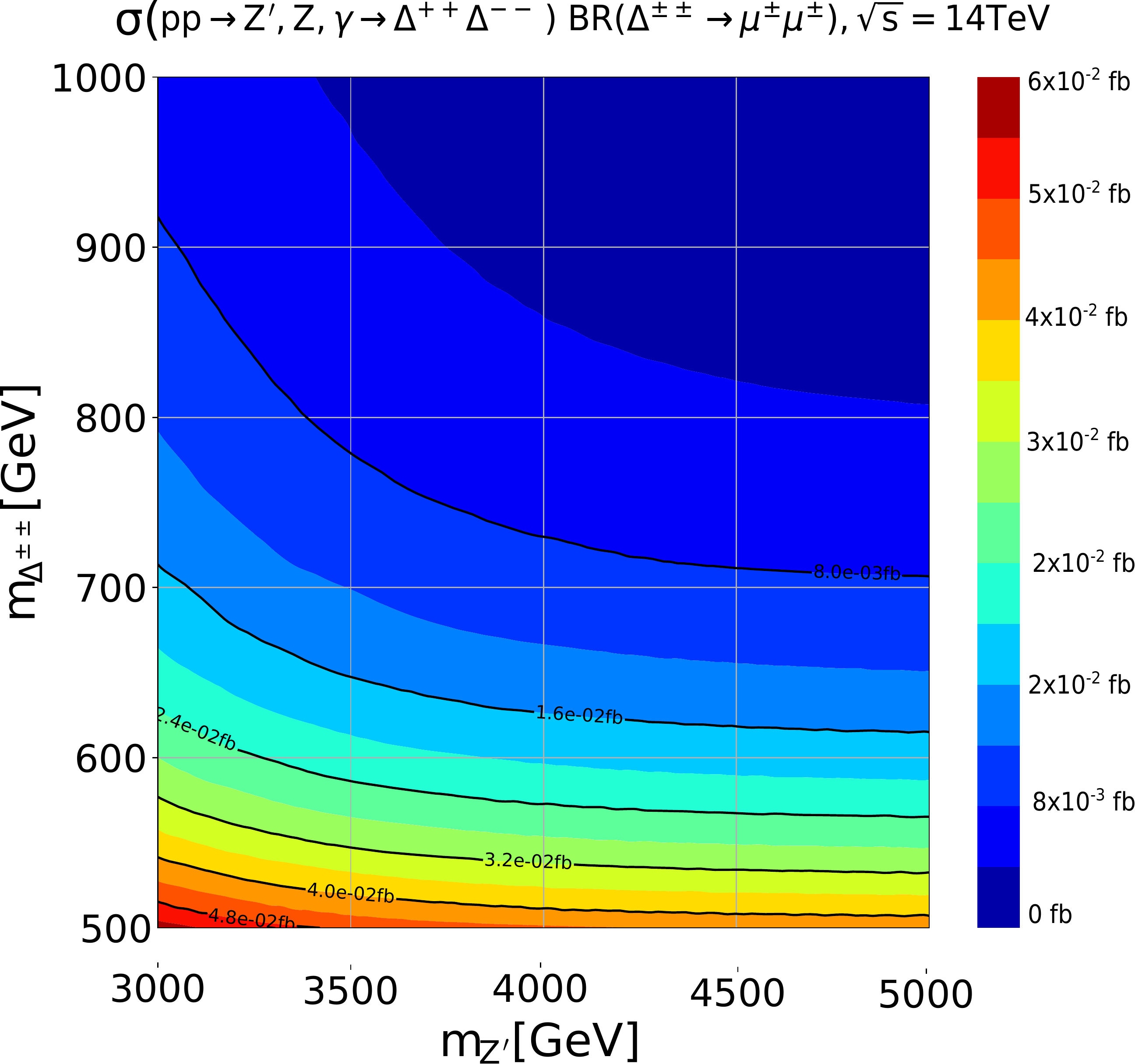}\,\,\,\includegraphics[scale=0.22]{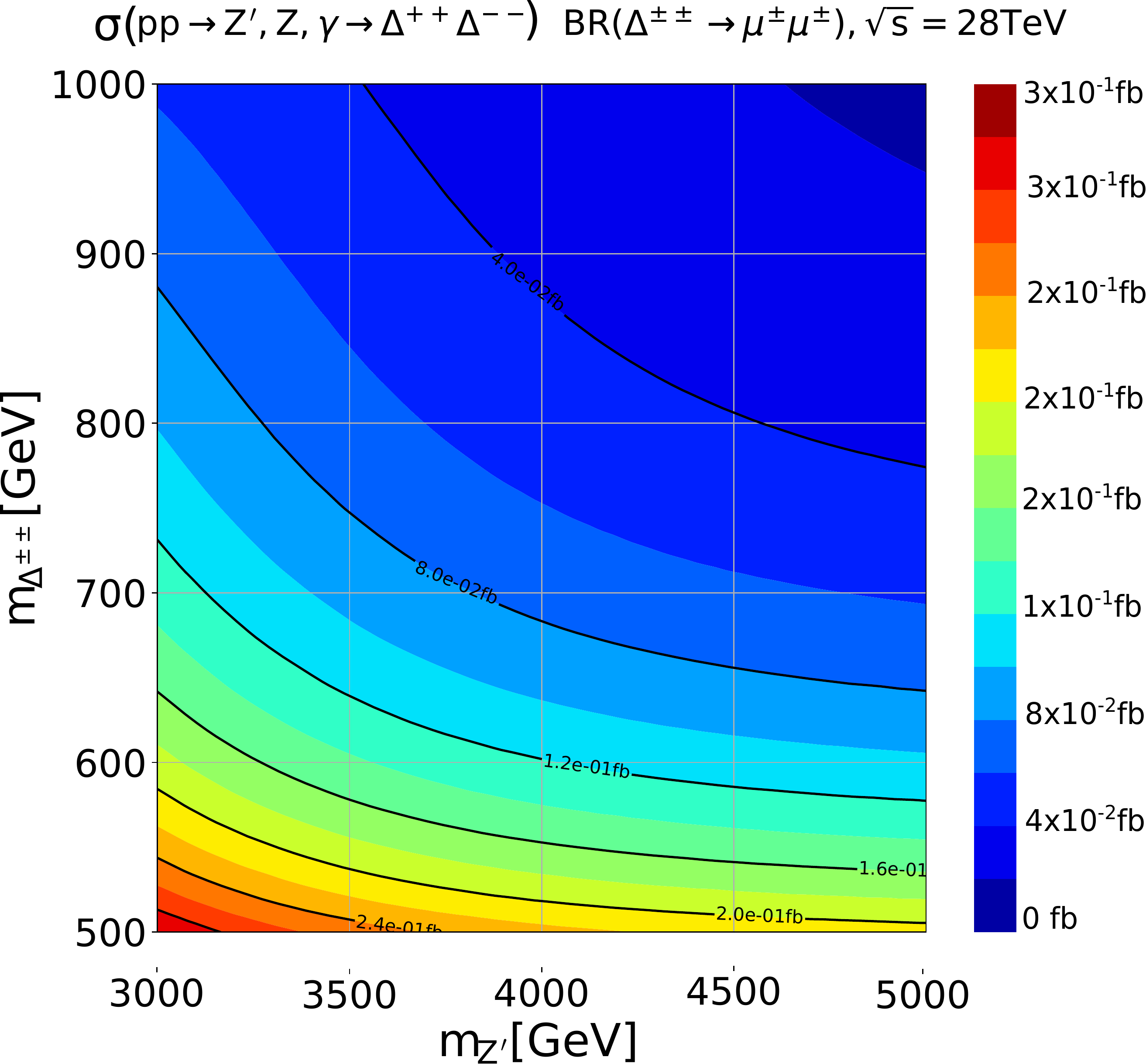}\,\,\,\includegraphics[scale=0.22]{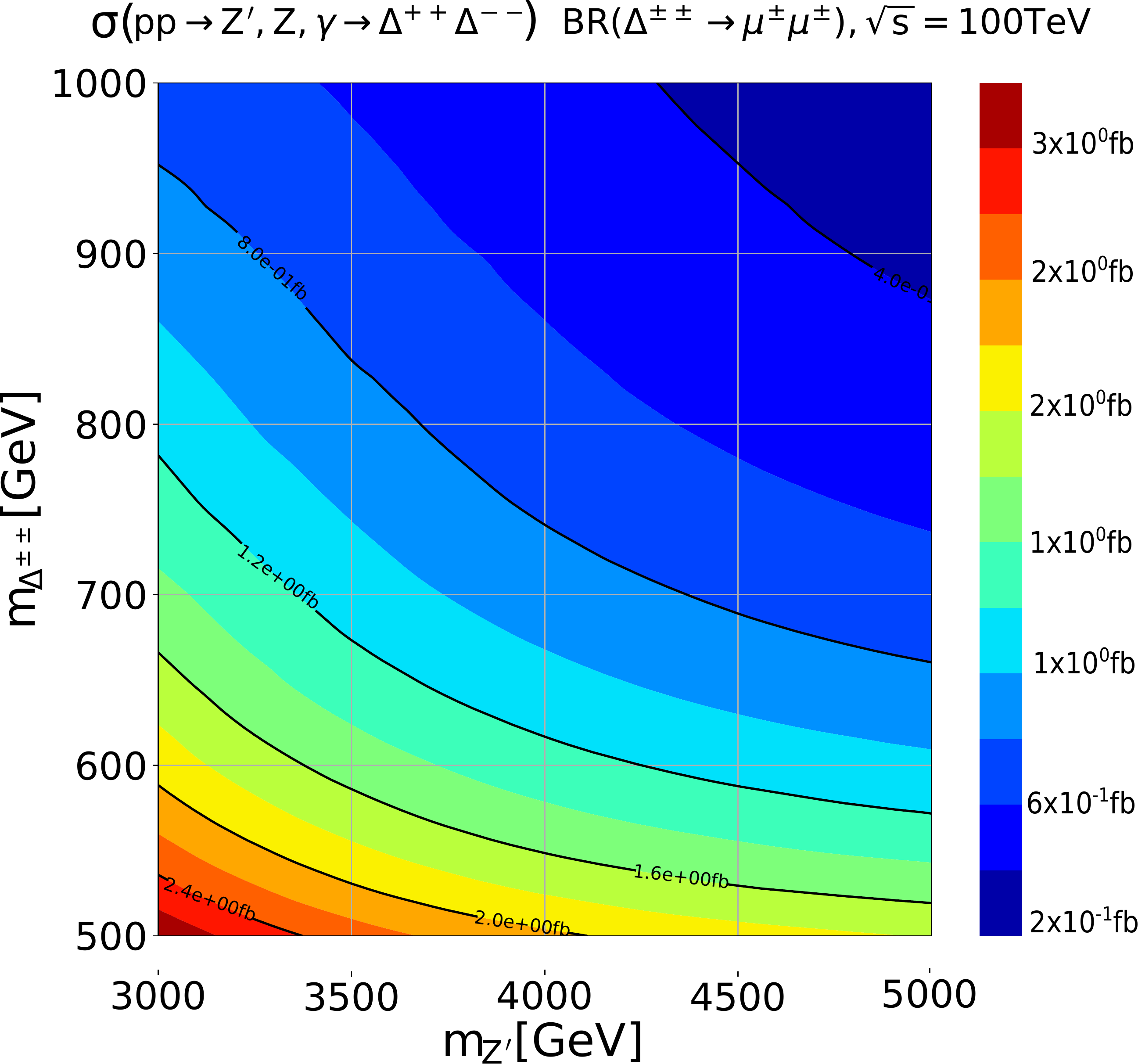}
\caption{Behavior of the cross section $\sigma (p\,p \rightarrow Z^{\prime} \rightarrow \Delta^{++}\,\Delta^{--} \rightarrow \mu^{+}\mu^{+}\mu^{-}\mu^{-})$ with $m_{Z^\prime}$ and $m_{\Delta^{\pm\pm}}$ varying at LHC with 14\,TeV, 28\,TeV and 100\,TeV, respectively.}
\label{Xsection_production_2Delta_4mu}
\end{figure}

\begin{figure}[ht]
\centering
\includegraphics[scale=0.22]{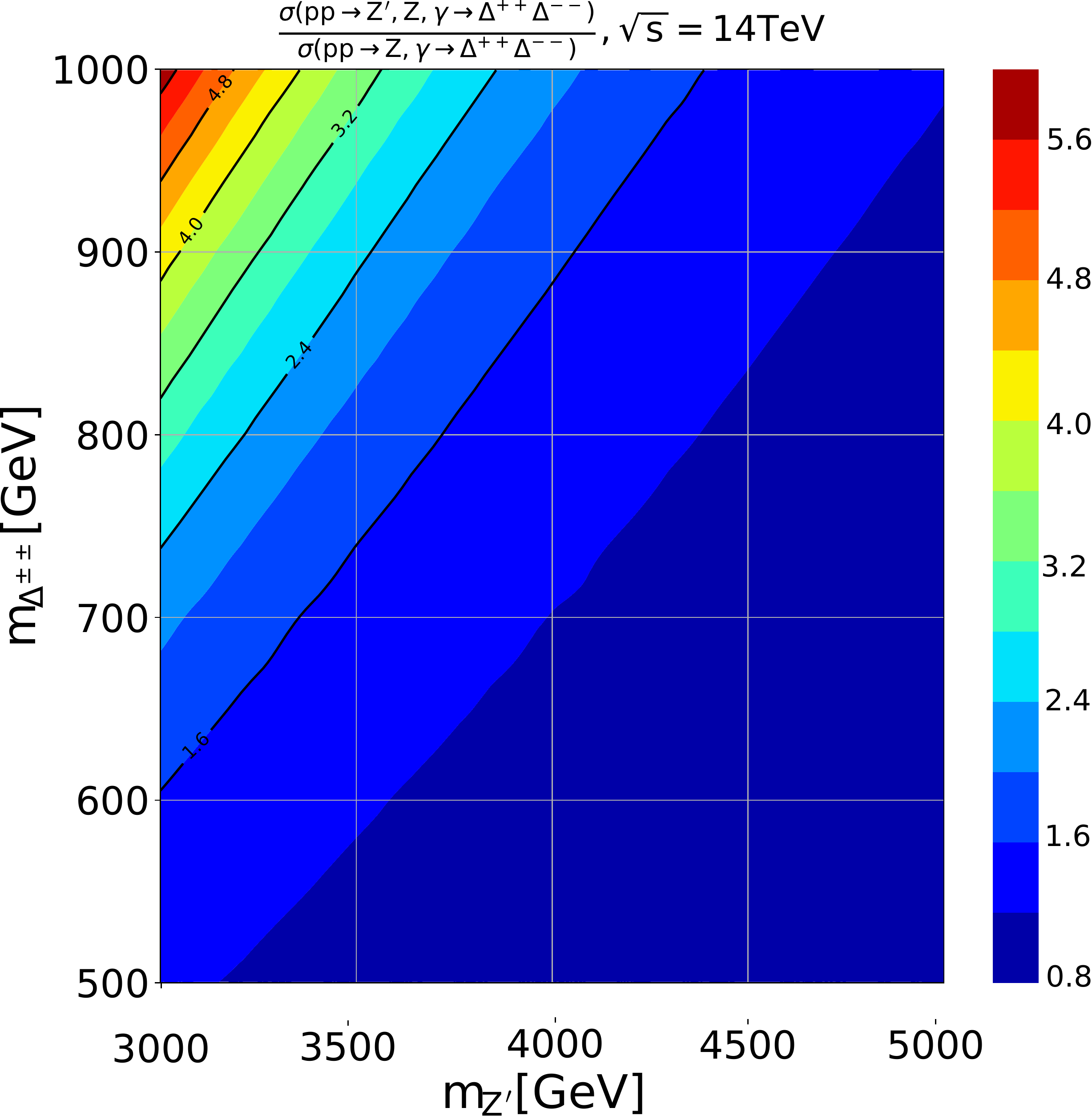}\,\,\,\includegraphics[scale=0.22]{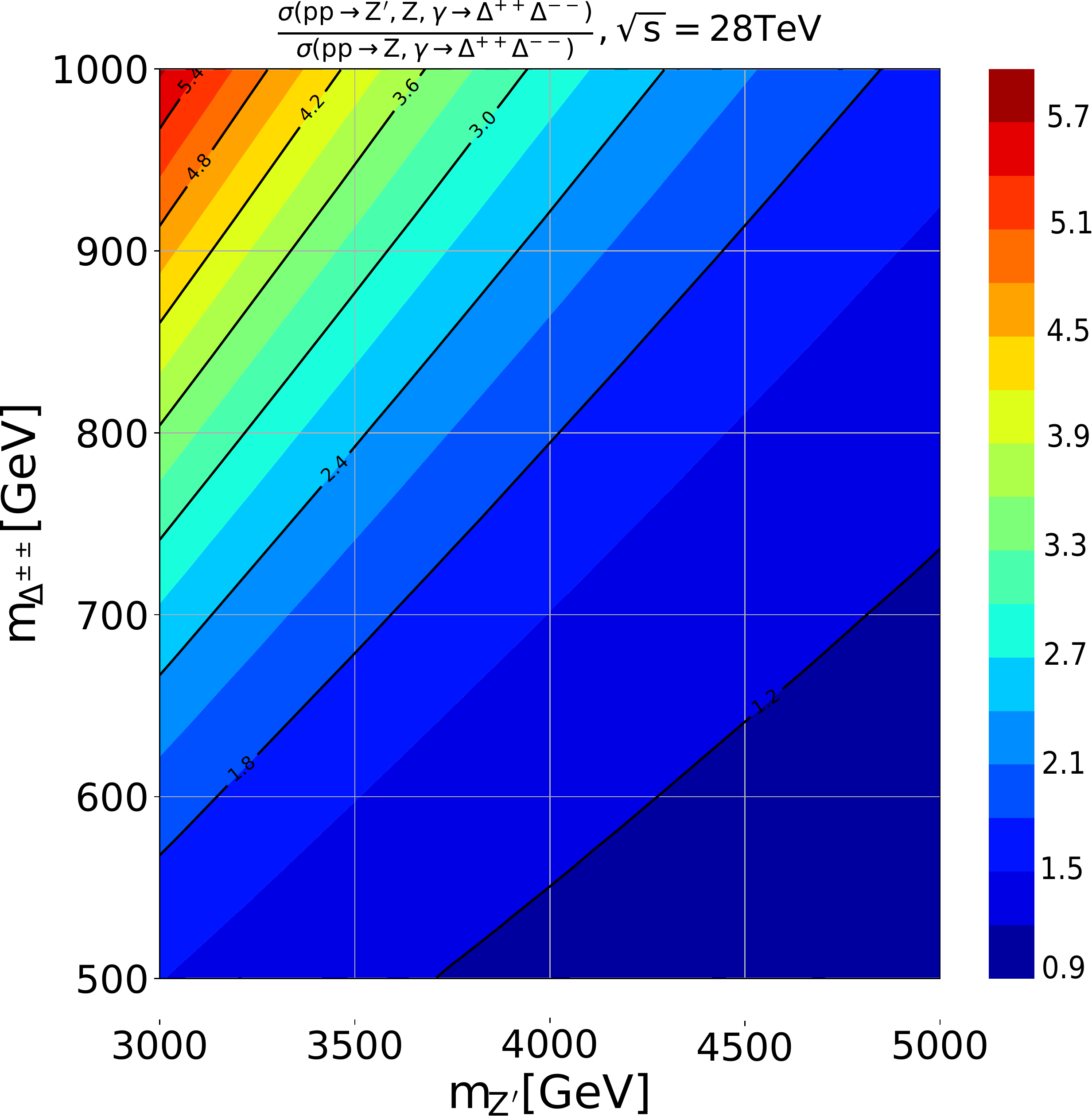}\,\,\,\includegraphics[scale=0.22]{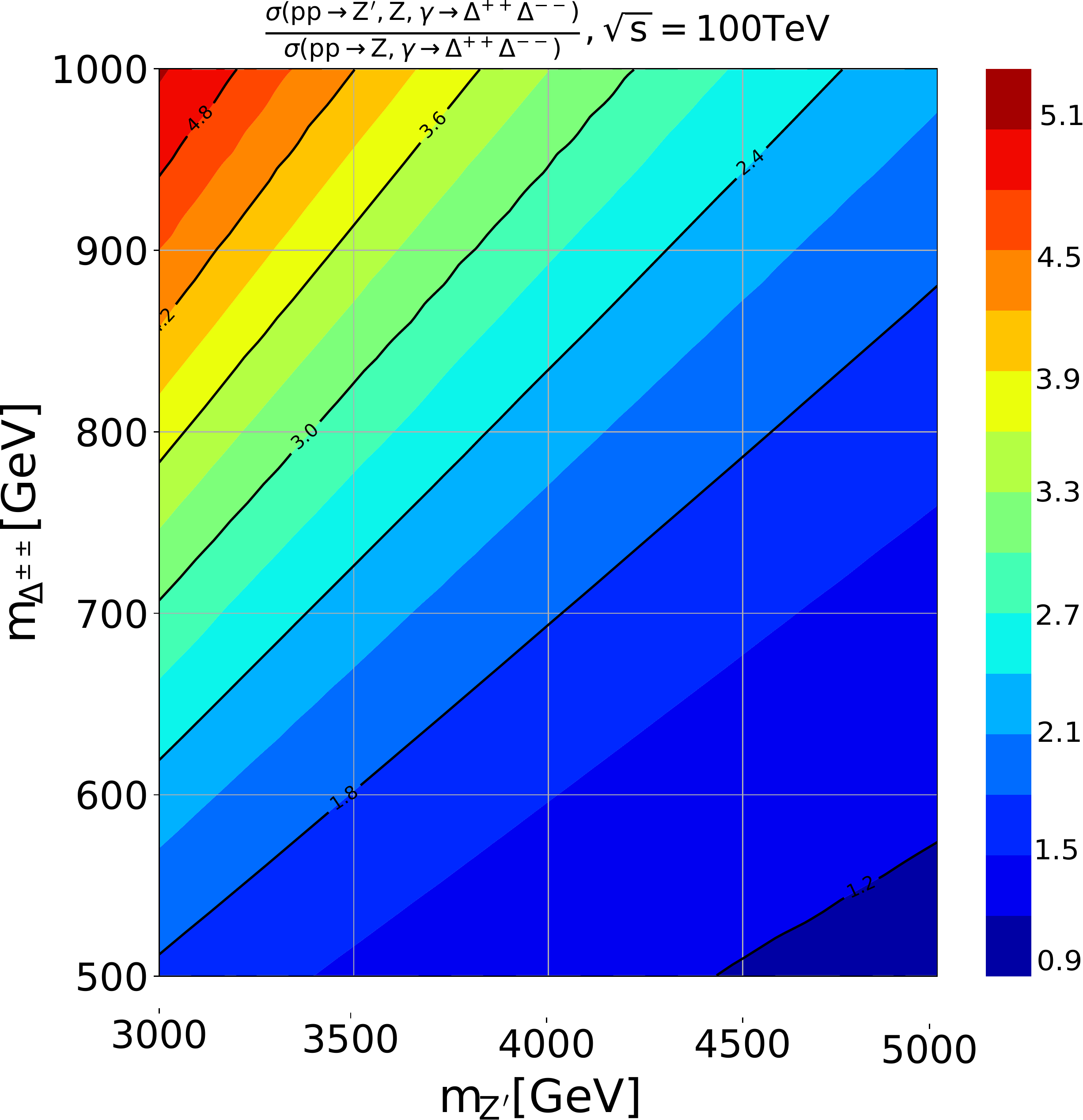}
\caption{ Cross section ratio $\sigma(p\,p \rightarrow Z^\prime, Z, \gamma \rightarrow \Delta^{\pm\pm}\,\Delta^{\pm\pm})/\sigma(p\,p \rightarrow Z,\gamma \rightarrow \Delta^{\pm\pm}\,\Delta^{\pm\pm}) $ at 14\,TeV, 28\,TeV and 100\,TeV. }
\label{R_production_4mu}
\end{figure}

The results of our analysis are displayed in Figs.\ref{Xsection_production_2Delta_4mu} and \ref{R_production_4mu}. In Fig.\ref{Xsection_production_2Delta_4mu}, we present the cross section  dependence with the masses of $\Delta^{++}$ and $Z^{\prime}$ for the LHC running with   energy of 14\,TeV, 28\,TeV and 100\,TeV. As we mentioned before, the chosen set of Yukawa couplings are not robust enough to provide sufficient events for the $\sqrt{s}=14$\,TeV center of mass energy. However, for the LHC running at 28\,TeV or 100\,TeV, our result are optimistic regarding the value of the cross section and consequently the expected number of events. We emphasize that these results are not definitive once it is plausible to enhance the Yukawa couplings by handling fairly values to the VEVs of the sextet and consequently improving the results of the processes. 

We remark that the contributions from the $Z^{\prime}$ neutral gauge bosons to the $\Delta^{\pm\pm}$ production  play the key role in the detection of the same. We verify this by estimating the ratio $\frac{\sigma (p\,p \rightarrow Z, \gamma ,Z^{\prime} \rightarrow \Delta^{++}\,\Delta^{--}) }{\sigma (p\,p \rightarrow Z, \gamma  \rightarrow \Delta^{++}\,\Delta^{--}) }$. In Fig. \ref{R_production_4mu} we present the behavior of this ratio with the mass of $\Delta^{++}$ and $Z^{\prime}$. As one can see, the contributions from the $Z^{\prime}$ neutral gauge bosons drastically modify the cross section of the full process and turns the signature viable for detection at future LHC runs .
%%%%

%%%

We now turn our attention to the reconstruction and identification of $Z^{\prime}$ and $\Delta^{\pm\pm}$. For that we make use of the profile likelihood for the process $ p\,p \rightarrow Z^{\prime} \rightarrow \Delta^{++}\,\Delta^{--} \rightarrow \mu^{+}\mu^{+}\mu^{-}\mu^{-}$. We generate  50 thousand events for the signal and 450 thousand events for the background in MadGraph. The background process can be generalized as the process $ p\,p \rightarrow \mu^{+}\mu^{+}\mu^{-}\mu^{-}$. For the acceptance criteria, we require the presence of at least four muons in the final state with $p_{T} > 32$GeV and $\vert\eta\vert < 2.1$. The relative muon isolation, the sum of transverse momenta of other particles in a cone of size $\Delta R = \sqrt[]{(\Delta\phi)^{2} + (\Delta\eta)^{2}} = 0.4$ around the direction of the candidate muon divided by the muon transverse momentum, is required to be less than 0.2. For the profile likelihood we keep the values of $M_{Z'}=4$TeV and $M_{\Delta}=700$GeV.

\begin{figure}[ht]
\centering
\includegraphics[scale=0.5]{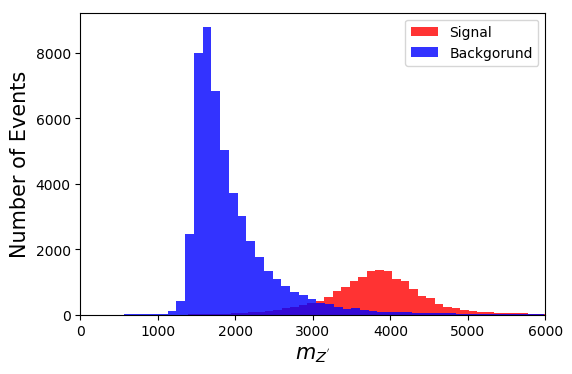}
\caption{Reconstructed invariant mass $M(\mu^{+},\mu^{-},\mu^{+},\mu^{-})$ for signal and background in the 4 muon channel.}
\label{MZPres}
\end{figure}

We fully reconstruct $Z^{\prime}$ using the object selection $M(\mu^{+},\mu^{-},\mu^{+},\mu^{-})$ which takes the mass of the four muons at the final state. In the fig. \ref{MZPres} we show the plot for $M(\mu^{+},\mu^{-},\mu^{+},\mu^{-})$ for signal and background events.

As one can see, the majority of the events in the background remains below 1.5 TeV for the $M(\mu^{+},\mu^{-},\mu^{+},\mu^{-})$ object. Then we can safely conclude that the background contribution to the region we are interest to probe is below 0.05\%.

With this in mind, we estimate the profile likelihood for identification of the double charged scalar and the $Z^{\prime}$ over the background. Our results are displayed at Fig. 5. We conclude that $m_{Z'}$ and $m_{\Delta}$ can be measured with very good precision, of course that the systematic errors, including pdf uncertainties and the detector energy resolution, will be dominant for this case but their contributions are expected to not exceed few \%. 

\begin{figure}[ht]
\centering
\includegraphics[scale=0.5]{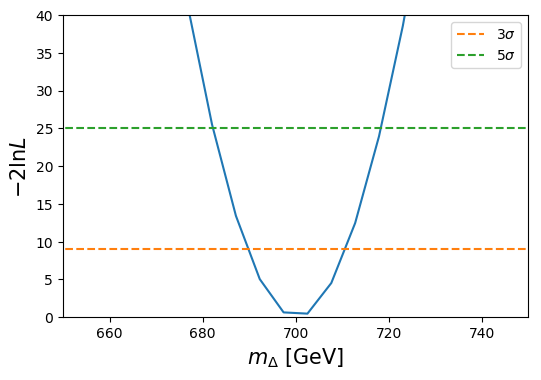}\,\,\,\includegraphics[scale=0.5]{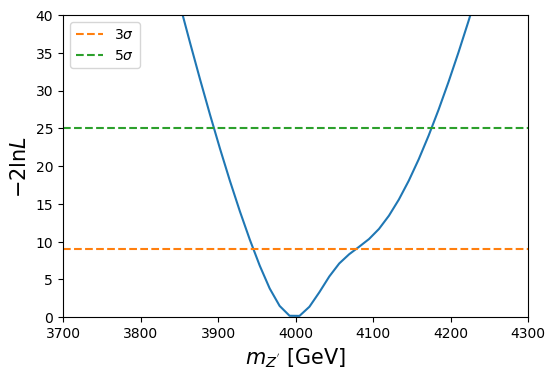}
\caption{ Likelihood profile for the masses of $\Delta$ (left) and Z' (right) measured using the fully reconstructed Z' in the channel $pp\rightarrow Z^{'}\rightarrow \Delta^{++}\Delta^{--}\rightarrow \mu^{+},\mu^{+},\mu^{-},\mu^{-}$.}
\label{likelihoos}
\end{figure}

\section{Conclusions}
\label{sec_V}
The inverse type II seesaw mechanism is a phenomenologically viable seesaw  mechanism because its signature ought to manifests at TeV scale and, in this way,  it can be probed at the LHC.   When adapted to the 3-3-1 model with right-handed neutrinos, the mechanism has the capacity of  generating small masses to both left-handed and right-handed neutrinos. In this point we draw attention to two facts. First, in the mechanism developed here the right-handed neutrinos are truly sterile neutrinos. This was obtained by assuming that the neutral component of the sextet, $\Phi^0$, does not develop VEV different from zero. Consequently, they do not  interact with the standard gauge bosons. This fact avoids all the current cosmological constraints on light right-handed neutrinos that arise from the very early universe\cite{Dolgov:2000jw,Ruchayskiy:2011aa,Vincent:2014rja}. On the other hand, they are active in relation to interactions with the gauge bosons characteristic of the 3-3-1 symmetry, namely $V^\pm$, $Z^{\prime}$ and $U^0$.  Second, the sextet scalar content decouples from the original scalar content  of the 3-3-1 model. This facilitates the search for the signature of the model in the form of charged scalars that compose the sextet. In our study we restricted our investigation to the sought for the doubly charged scalars by producing it at the LHC through the process  $\sigma (p\,p \rightarrow Z, \gamma ,Z^{\prime} \rightarrow \Delta^{++}\,\Delta^{--}) $ followed by their decays in pairs of muons and anti-muons. Our results suggest that both $Z^{\prime}$ and the double charged scalars  maybe be detected at the LHC with $\sqrt{s}=28$\,TeV, but the higher chance to probe these new particles remains in the future $\sqrt{s}=100$\,TeV collider.
%%%%%

\section*{Acknowledgments}
C.A.S.P  was supported by the CNPq research grants, No. 304423/2017-3,  P.V thanks  Coordena\c c\~ ao de Aperfeicoamento de Pessoal de N\'ivel Superior - CAPES, F.F.F s supported in part by project Y8Y2411B11, China Postdoctoral Science Fundation. L.H. is supported in part by the US National Science Foundation grant PHY-1719877. J.S. is supported by the NSFC under grant No.11647601, No.11690022, No.11851302, No.11675243 and No.11761141011 and also supported by the Strategic Priority Research Program of the Chinese Academy of Sciences under grant No.XDB21010200 and No.XDB23000000. The simulations for this work were done in part at the HPC Cluster of ITP-CAS.

%%%%%

\bibliographystyle{JHEPfixed}
\bibliography{sample}

\begin{thebibliography}{31}
\expandafter\ifx\csname natexlab\endcsname\relax\def\natexlab#1{#1}\fi
\expandafter\ifx\csname bibnamefont\endcsname\relax
  \def\bibnamefont#1{#1}\fi
\expandafter\ifx\csname bibfnamefont\endcsname\relax
  \def\bibfnamefont#1{#1}\fi
\expandafter\ifx\csname citenamefont\endcsname\relax
  \def\citenamefont#1{#1}\fi
\expandafter\ifx\csname url\endcsname\relax
  \def\url#1{\texttt{#1}}\fi
\expandafter\ifx\csname urlprefix\endcsname\relax\def\urlprefix{URL }\fi
\providecommand{\bibinfo}[2]{#2}
\providecommand{\eprint}[2][]{\url{#2}}

\bibitem[{\citenamefont{Mohapatra and Valle}(1986)}]{Mohapatra:1986bd}
\bibinfo{author}{\bibfnamefont{R.~N.} \bibnamefont{Mohapatra}}
  \bibnamefont{and} \bibinfo{author}{\bibfnamefont{J.~W.~F.}
  \bibnamefont{Valle}}, \bibinfo{journal}{Phys. Rev.}
  \textbf{\bibinfo{volume}{D34}}, \bibinfo{pages}{1642} (\bibinfo{year}{1986}),
  \bibinfo{note}{[,235(1986)]}.

\bibitem[{\citenamefont{Li et~al.}(1985)\citenamefont{Li, Liu, and
  Wolfenstein}}]{Li:1985hy}
\bibinfo{author}{\bibfnamefont{L.-F.} \bibnamefont{Li}},
  \bibinfo{author}{\bibfnamefont{Y.}~\bibnamefont{Liu}}, \bibnamefont{and}
  \bibinfo{author}{\bibfnamefont{L.}~\bibnamefont{Wolfenstein}},
  \bibinfo{journal}{Phys. Lett.} \textbf{\bibinfo{volume}{159B}},
  \bibinfo{pages}{45} (\bibinfo{year}{1985}).

\bibitem[{\citenamefont{Lusignoli et~al.}(1990)\citenamefont{Lusignoli,
  Masiero, and Roncadelli}}]{Lusignoli:1990yk}
\bibinfo{author}{\bibfnamefont{M.}~\bibnamefont{Lusignoli}},
  \bibinfo{author}{\bibfnamefont{A.}~\bibnamefont{Masiero}}, \bibnamefont{and}
  \bibinfo{author}{\bibfnamefont{M.}~\bibnamefont{Roncadelli}},
  \bibinfo{journal}{Phys. Lett.} \textbf{\bibinfo{volume}{B252}},
  \bibinfo{pages}{247} (\bibinfo{year}{1990}).

\bibitem[{\citenamefont{de~S.~Pires}(2006)}]{deS.Pires:2005au}
\bibinfo{author}{\bibfnamefont{C.~A.} \bibnamefont{de~S.~Pires}},
  \bibinfo{journal}{Mod. Phys. Lett.} \textbf{\bibinfo{volume}{A21}},
  \bibinfo{pages}{971} (\bibinfo{year}{2006}), \eprint{hep-ph/0509152}.

\bibitem[{\citenamefont{Freitas et~al.}(2017)\citenamefont{Freitas,
  de~S.~Pires, and Rodrigues~da Silva}}]{Freitas:2014fda}
\bibinfo{author}{\bibfnamefont{F.~F.} \bibnamefont{Freitas}},
  \bibinfo{author}{\bibfnamefont{C.~A.} \bibnamefont{de~S.~Pires}},
  \bibnamefont{and} \bibinfo{author}{\bibfnamefont{P.~S.}
  \bibnamefont{Rodrigues~da Silva}}, \bibinfo{journal}{Phys. Lett.}
  \textbf{\bibinfo{volume}{B769}}, \bibinfo{pages}{48} (\bibinfo{year}{2017}),
  \eprint{1408.5878}.

\bibitem[{\citenamefont{Ma}(2009)}]{Ma:2009kh}
\bibinfo{author}{\bibfnamefont{E.}~\bibnamefont{Ma}}, \bibinfo{journal}{Mod.
  Phys. Lett.} \textbf{\bibinfo{volume}{A24}}, \bibinfo{pages}{2491}
  (\bibinfo{year}{2009}), \eprint{0905.2972}.

\bibitem[{\citenamefont{Ibanez et~al.}(2009)\citenamefont{Ibanez, Morisi, and
  Valle}}]{Ibanez:2009du}
\bibinfo{author}{\bibfnamefont{D.}~\bibnamefont{Ibanez}},
  \bibinfo{author}{\bibfnamefont{S.}~\bibnamefont{Morisi}}, \bibnamefont{and}
  \bibinfo{author}{\bibfnamefont{J.~W.~F.} \bibnamefont{Valle}},
  \bibinfo{journal}{Phys. Rev.} \textbf{\bibinfo{volume}{D80}},
  \bibinfo{pages}{053015} (\bibinfo{year}{2009}), \eprint{0907.3109}.

\bibitem[{\citenamefont{Singer et~al.}(1980)\citenamefont{Singer, Valle, and
  Schechter}}]{Singer:1980sw}
\bibinfo{author}{\bibfnamefont{M.}~\bibnamefont{Singer}},
  \bibinfo{author}{\bibfnamefont{J.~W.~F.} \bibnamefont{Valle}},
  \bibnamefont{and}
  \bibinfo{author}{\bibfnamefont{J.}~\bibnamefont{Schechter}},
  \bibinfo{journal}{Phys. Rev.} \textbf{\bibinfo{volume}{D22}},
  \bibinfo{pages}{738} (\bibinfo{year}{1980}).

\bibitem[{\citenamefont{Montero et~al.}(1993)\citenamefont{Montero, Pisano, and
  Pleitez}}]{Montero:1992jk}
\bibinfo{author}{\bibfnamefont{J.~C.} \bibnamefont{Montero}},
  \bibinfo{author}{\bibfnamefont{F.}~\bibnamefont{Pisano}}, \bibnamefont{and}
  \bibinfo{author}{\bibfnamefont{V.}~\bibnamefont{Pleitez}},
  \bibinfo{journal}{Phys. Rev.} \textbf{\bibinfo{volume}{D47}},
  \bibinfo{pages}{2918} (\bibinfo{year}{1993}), \eprint{hep-ph/9212271}.

\bibitem[{\citenamefont{Foot et~al.}(1994)\citenamefont{Foot, Long, and
  Tran}}]{Foot:1994ym}
\bibinfo{author}{\bibfnamefont{R.}~\bibnamefont{Foot}},
  \bibinfo{author}{\bibfnamefont{H.~N.} \bibnamefont{Long}}, \bibnamefont{and}
  \bibinfo{author}{\bibfnamefont{T.~A.} \bibnamefont{Tran}},
  \bibinfo{journal}{Phys. Rev.} \textbf{\bibinfo{volume}{D50}},
  \bibinfo{pages}{R34} (\bibinfo{year}{1994}), \eprint{hep-ph/9402243}.

\bibitem[{\citenamefont{Hettmansperger
  et~al.}(2011)\citenamefont{Hettmansperger, Lindner, and
  Rodejohann}}]{Hettmansperger:2011bt}
\bibinfo{author}{\bibfnamefont{H.}~\bibnamefont{Hettmansperger}},
  \bibinfo{author}{\bibfnamefont{M.}~\bibnamefont{Lindner}}, \bibnamefont{and}
  \bibinfo{author}{\bibfnamefont{W.}~\bibnamefont{Rodejohann}},
  \bibinfo{journal}{JHEP} \textbf{\bibinfo{volume}{04}}, \bibinfo{pages}{123}
  (\bibinfo{year}{2011}), \eprint{1102.3432}.

\bibitem[{\citenamefont{Minkowski}(1977)}]{Minkowski:1977sc}
\bibinfo{author}{\bibfnamefont{P.}~\bibnamefont{Minkowski}},
  \bibinfo{journal}{Phys. Lett.} \textbf{\bibinfo{volume}{67B}},
  \bibinfo{pages}{421} (\bibinfo{year}{1977}).

\bibitem[{\citenamefont{Yanagida}(1979)}]{Yanagida:1979as}
\bibinfo{author}{\bibfnamefont{T.}~\bibnamefont{Yanagida}},
  \bibinfo{journal}{Conf. Proc.} \textbf{\bibinfo{volume}{C7902131}},
  \bibinfo{pages}{95} (\bibinfo{year}{1979}).

\bibitem[{\citenamefont{Gell-Mann et~al.}(1979)\citenamefont{Gell-Mann, Ramond,
  and Slansky}}]{GellMann:1980vs}
\bibinfo{author}{\bibfnamefont{M.}~\bibnamefont{Gell-Mann}},
  \bibinfo{author}{\bibfnamefont{P.}~\bibnamefont{Ramond}}, \bibnamefont{and}
  \bibinfo{author}{\bibfnamefont{R.}~\bibnamefont{Slansky}},
  \bibinfo{journal}{Conf. Proc.} \textbf{\bibinfo{volume}{C790927}},
  \bibinfo{pages}{315} (\bibinfo{year}{1979}), \eprint{1306.4669}.

\bibitem[{\citenamefont{Mohapatra and Senjanovic}(1980)}]{Mohapatra:1979ia}
\bibinfo{author}{\bibfnamefont{R.~N.} \bibnamefont{Mohapatra}}
  \bibnamefont{and}
  \bibinfo{author}{\bibfnamefont{G.}~\bibnamefont{Senjanovic}},
  \bibinfo{journal}{Phys. Rev. Lett.} \textbf{\bibinfo{volume}{44}},
  \bibinfo{pages}{912} (\bibinfo{year}{1980}), \bibinfo{note}{[,231(1979)]}.

\bibitem[{\citenamefont{Dias et~al.}(2011)\citenamefont{Dias, de~S.~Pires, and
  da~Silva}}]{Dias:2011sq}
\bibinfo{author}{\bibfnamefont{A.~G.} \bibnamefont{Dias}},
  \bibinfo{author}{\bibfnamefont{C.~A.} \bibnamefont{de~S.~Pires}},
  \bibnamefont{and} \bibinfo{author}{\bibfnamefont{P.~S.~R.}
  \bibnamefont{da~Silva}}, \bibinfo{journal}{Phys. Rev.}
  \textbf{\bibinfo{volume}{D84}}, \bibinfo{pages}{053011}
  (\bibinfo{year}{2011}), \eprint{1107.0739}.

\bibitem[{\citenamefont{Das and Okada}(2013)}]{Das:2012ze}
\bibinfo{author}{\bibfnamefont{A.}~\bibnamefont{Das}} \bibnamefont{and}
  \bibinfo{author}{\bibfnamefont{N.}~\bibnamefont{Okada}},
  \bibinfo{journal}{Phys. Rev.} \textbf{\bibinfo{volume}{D88}},
  \bibinfo{pages}{113001} (\bibinfo{year}{2013}), \eprint{1207.3734}.

\bibitem[{\citenamefont{Das et~al.}(2014)\citenamefont{Das, Bhupal~Dev, and
  Okada}}]{Das:2014jxa}
\bibinfo{author}{\bibfnamefont{A.}~\bibnamefont{Das}},
  \bibinfo{author}{\bibfnamefont{P.~S.} \bibnamefont{Bhupal~Dev}},
  \bibnamefont{and} \bibinfo{author}{\bibfnamefont{N.}~\bibnamefont{Okada}},
  \bibinfo{journal}{Phys. Lett.} \textbf{\bibinfo{volume}{B735}},
  \bibinfo{pages}{364} (\bibinfo{year}{2014}), \eprint{1405.0177}.

\bibitem[{\citenamefont{Das et~al.}(2017{\natexlab{a}})\citenamefont{Das, Dev,
  and Kim}}]{Das:2017zjc}
\bibinfo{author}{\bibfnamefont{A.}~\bibnamefont{Das}},
  \bibinfo{author}{\bibfnamefont{P.~S.~B.} \bibnamefont{Dev}},
  \bibnamefont{and} \bibinfo{author}{\bibfnamefont{C.~S.} \bibnamefont{Kim}},
  \bibinfo{journal}{Phys. Rev.} \textbf{\bibinfo{volume}{D95}},
  \bibinfo{pages}{115013} (\bibinfo{year}{2017}{\natexlab{a}}),
  \eprint{1704.00880}.

\bibitem[{\citenamefont{Das et~al.}(2017{\natexlab{b}})\citenamefont{Das, Gao,
  and Kamon}}]{Das:2017rsu}
\bibinfo{author}{\bibfnamefont{A.}~\bibnamefont{Das}},
  \bibinfo{author}{\bibfnamefont{Y.}~\bibnamefont{Gao}}, \bibnamefont{and}
  \bibinfo{author}{\bibfnamefont{T.}~\bibnamefont{Kamon}}
  (\bibinfo{year}{2017}{\natexlab{b}}), \eprint{1704.00881}.

\bibitem[{\citenamefont{Bhupal~Dev et~al.}(2012)\citenamefont{Bhupal~Dev,
  Franceschini, and Mohapatra}}]{BhupalDev:2012zg}
\bibinfo{author}{\bibfnamefont{P.~S.} \bibnamefont{Bhupal~Dev}},
  \bibinfo{author}{\bibfnamefont{R.}~\bibnamefont{Franceschini}},
  \bibnamefont{and} \bibinfo{author}{\bibfnamefont{R.~N.}
  \bibnamefont{Mohapatra}}, \bibinfo{journal}{Phys. Rev.}
  \textbf{\bibinfo{volume}{D86}}, \bibinfo{pages}{093010}
  (\bibinfo{year}{2012}), \eprint{1207.2756}.

\bibitem[{\citenamefont{'t~Hooft}(1980)}]{tHooft:1979rat}
\bibinfo{author}{\bibfnamefont{G.}~\bibnamefont{'t~Hooft}},
  \bibinfo{journal}{NATO Sci. Ser. B} \textbf{\bibinfo{volume}{59}},
  \bibinfo{pages}{135} (\bibinfo{year}{1980}).

\bibitem[{\citenamefont{Dias et~al.}(2012)\citenamefont{Dias, de~S.~Pires,
  Rodrigues~da Silva, and Sampieri}}]{Dias:2012xp}
\bibinfo{author}{\bibfnamefont{A.~G.} \bibnamefont{Dias}},
  \bibinfo{author}{\bibfnamefont{C.~A.} \bibnamefont{de~S.~Pires}},
  \bibinfo{author}{\bibfnamefont{P.~S.} \bibnamefont{Rodrigues~da Silva}},
  \bibnamefont{and} \bibinfo{author}{\bibfnamefont{A.}~\bibnamefont{Sampieri}},
  \bibinfo{journal}{Phys. Rev.} \textbf{\bibinfo{volume}{D86}},
  \bibinfo{pages}{035007} (\bibinfo{year}{2012}), \eprint{1206.2590}.

\bibitem[{\citenamefont{Tanabashi et~al.}(2018)}]{Tanabashi:2018oca}
\bibinfo{author}{\bibfnamefont{M.}~\bibnamefont{Tanabashi}}
  \bibnamefont{et~al.} (\bibinfo{collaboration}{Particle Data Group}),
  \bibinfo{journal}{Phys. Rev.} \textbf{\bibinfo{volume}{D98}},
  \bibinfo{pages}{030001} (\bibinfo{year}{2018}).

\bibitem[{\citenamefont{Long et~al.}(2016)\citenamefont{Long, Hue, and
  Loi}}]{Long:2016lmj}
\bibinfo{author}{\bibfnamefont{H.~N.} \bibnamefont{Long}},
  \bibinfo{author}{\bibfnamefont{L.~T.} \bibnamefont{Hue}}, \bibnamefont{and}
  \bibinfo{author}{\bibfnamefont{D.~V.} \bibnamefont{Loi}},
  \bibinfo{journal}{Phys. Rev.} \textbf{\bibinfo{volume}{D94}},
  \bibinfo{pages}{015007} (\bibinfo{year}{2016}), \eprint{1605.07835}.

\bibitem[{\citenamefont{Alloul et~al.}(2014)\citenamefont{Alloul, Christensen,
  Degrande, Duhr, and Fuks}}]{Alloul:2013bka}
\bibinfo{author}{\bibfnamefont{A.}~\bibnamefont{Alloul}},
  \bibinfo{author}{\bibfnamefont{N.~D.} \bibnamefont{Christensen}},
  \bibinfo{author}{\bibfnamefont{C.}~\bibnamefont{Degrande}},
  \bibinfo{author}{\bibfnamefont{C.}~\bibnamefont{Duhr}}, \bibnamefont{and}
  \bibinfo{author}{\bibfnamefont{B.}~\bibnamefont{Fuks}},
  \bibinfo{journal}{Comput. Phys. Commun.} \textbf{\bibinfo{volume}{185}},
  \bibinfo{pages}{2250} (\bibinfo{year}{2014}), \eprint{1310.1921}.

\bibitem[{\citenamefont{Degrande et~al.}(2012)\citenamefont{Degrande, Duhr,
  Fuks, Grellscheid, Mattelaer, and Reiter}}]{Degrande:2011ua}
\bibinfo{author}{\bibfnamefont{C.}~\bibnamefont{Degrande}},
  \bibinfo{author}{\bibfnamefont{C.}~\bibnamefont{Duhr}},
  \bibinfo{author}{\bibfnamefont{B.}~\bibnamefont{Fuks}},
  \bibinfo{author}{\bibfnamefont{D.}~\bibnamefont{Grellscheid}},
  \bibinfo{author}{\bibfnamefont{O.}~\bibnamefont{Mattelaer}},
  \bibnamefont{and} \bibinfo{author}{\bibfnamefont{T.}~\bibnamefont{Reiter}},
  \bibinfo{journal}{Comput. Phys. Commun.} \textbf{\bibinfo{volume}{183}},
  \bibinfo{pages}{1201} (\bibinfo{year}{2012}), \eprint{1108.2040}.

\bibitem[{\citenamefont{Fileviez~Perez
  et~al.}(2008)\citenamefont{Fileviez~Perez, Han, Huang, Li, and
  Wang}}]{Perez:2008ha}
\bibinfo{author}{\bibfnamefont{P.}~\bibnamefont{Fileviez~Perez}},
  \bibinfo{author}{\bibfnamefont{T.}~\bibnamefont{Han}},
  \bibinfo{author}{\bibfnamefont{G.-y.} \bibnamefont{Huang}},
  \bibinfo{author}{\bibfnamefont{T.}~\bibnamefont{Li}}, \bibnamefont{and}
  \bibinfo{author}{\bibfnamefont{K.}~\bibnamefont{Wang}},
  \bibinfo{journal}{Phys. Rev.} \textbf{\bibinfo{volume}{D78}},
  \bibinfo{pages}{015018} (\bibinfo{year}{2008}), \eprint{0805.3536}.

\bibitem[{\citenamefont{Dolgov et~al.}(2000)\citenamefont{Dolgov, Hansen,
  Raffelt, and Semikoz}}]{Dolgov:2000jw}
\bibinfo{author}{\bibfnamefont{A.~D.} \bibnamefont{Dolgov}},
  \bibinfo{author}{\bibfnamefont{S.~H.} \bibnamefont{Hansen}},
  \bibinfo{author}{\bibfnamefont{G.}~\bibnamefont{Raffelt}}, \bibnamefont{and}
  \bibinfo{author}{\bibfnamefont{D.~V.} \bibnamefont{Semikoz}},
  \bibinfo{journal}{Nucl. Phys.} \textbf{\bibinfo{volume}{B590}},
  \bibinfo{pages}{562} (\bibinfo{year}{2000}), \eprint{hep-ph/0008138}.

\bibitem[{\citenamefont{Ruchayskiy and Ivashko}(2012)}]{Ruchayskiy:2011aa}
\bibinfo{author}{\bibfnamefont{O.}~\bibnamefont{Ruchayskiy}} \bibnamefont{and}
  \bibinfo{author}{\bibfnamefont{A.}~\bibnamefont{Ivashko}},
  \bibinfo{journal}{JHEP} \textbf{\bibinfo{volume}{06}}, \bibinfo{pages}{100}
  (\bibinfo{year}{2012}), \eprint{1112.3319}.

\bibitem[{\citenamefont{Vincent et~al.}(2015)\citenamefont{Vincent, Martinez,
  Hernández, Lattanzi, and Mena}}]{Vincent:2014rja}
\bibinfo{author}{\bibfnamefont{A.~C.} \bibnamefont{Vincent}},
  \bibinfo{author}{\bibfnamefont{E.~F.} \bibnamefont{Martinez}},
  \bibinfo{author}{\bibfnamefont{P.}~\bibnamefont{Hernández}},
  \bibinfo{author}{\bibfnamefont{M.}~\bibnamefont{Lattanzi}}, \bibnamefont{and}
  \bibinfo{author}{\bibfnamefont{O.}~\bibnamefont{Mena}},
  \bibinfo{journal}{JCAP} \textbf{\bibinfo{volume}{1504}}, \bibinfo{pages}{006}
  (\bibinfo{year}{2015}), \eprint{1408.1956}.

\end{thebibliography}

\end{document}